Chapter 6

**"That star is not on the map": the German side of the discovery**

**Davor Krajnović**
Leibniz-Institut für Astrophysik Potsdam (AIP)
An der Sternwarte 16
14482 Potsdam, Germany
email: dkrajnovic@aip.de

**Abstract.** Neptune was telescopically discovered by Johan Gottfried Galle and Heinrich Louis d'Arrest in Berlin on 23 September 1846 based on the prediction by Urbain Jean Joseph Le Verrier. The role German astronomers played in the discovery has often been overshadowed by the controversies that erupted in England and France after the discovery. However, their role was crucial, not only in bringing about the discovery in the first place, but also in resolving some of the post-discovery controversies that erupted around priority for the prediction and naming of the planet. German astronomers in Central and Eastern Europe possessed some of the best telescopes of the day and had established themselves over several decades as being at the forefront of observational astronomy. They had produced the star charts that in the end proved indispensable for allowing identification of the planet, while a German publication, *Astronomische Nachrichten*, published by Heinrich Christian Schumacher in Altona, in the vicinity of Hamburg, then part of the Kingdom of Denmark, served as the journal of record during the time period with which these events took place. The general neglect of the German part of the story is most strongly attested by the inaccuracies concerning what actually happened on discovery night, which were long propagated in the English-speaking literature. Notably, for 30 years after the discovery, it was not appreciated that there were two observers, while the role of the sky map used was often exaggerated. This chapter sets forth a more complete picture, and in particular emphasises more than previous accounts the critical role of d'Arrest, arguing that he should be celebrated as a co-discoverer. In addition, evidence presented here, much of it not previously available in the English-speaking literature, shows that the name "Neptune" was eventually accepted throughout the scientific community based on German precedents.

**6.1 Introduction**

In his autobiography, written long after the discovery of Neptune, Astronomer Royal George Biddell Airy has only this to say about what he calls the "engrossing subject" of the year 1846.:

> … As I have said (1845) I obtained no answer from Adams to a letter of enquiry. Beginning with June 26th of 1846 I had correspondence of a satisfactory character with Le Verrier, who had taken up the subject of the disturbance of Uranus, and arrived at conclusions not very different from those of Adams. I wrote from Ely on July 9th to Challis, begging him, as in possession of the largest telescope in England, to sweep for the planet, and suggesting a plan. I received information of its recognition by Galle, when I was visiting Hansen at Gotha. For further official history, see my communications to the Royal Astronomical Society, and for private history see the papers of the Royal Observatory. I was abused most savagely both by the English and French. (Airy, 1896:181)

As Airy noted here, he was in Germany when he learned of the discovery of the planet by a German observer at Berlin. He singled out the English and the French as the nationalities that subjected him to savage abuse. Curiously, he did not mention Germans. As we have seen, the international rivalry that erupted after the discovery was between the French, who in Urbain Jean Joseph Le Verrier claimed the only mathematician who published a prediction, and the British, who claimed a prior, although unpublished, prediction by John Couch Adams. The British side also could add James Challis's seemingly rather secretive search at Cambridge in the summer before the planet was actually found (see Chap. 4 and 8).

   Despite having been responsible for the actual discovery, about which there was never any contest, the Germans were rather sidelined, and long rendered almost to the status of a footnote as the dispute raged between the two theoreticians and their respective supporters. Yet, the optical discovery clearly belonged to Germans and them alone. Whereas Adams's calculations could be rehabilitated for at least partial credit, there seemed nothing creditable in English search, which had ended in nearly total failure, underscored by the embarrassing failing to recognise the planet on two occasions before the definitive news from Berlin arrived. Thus, Airy's role, whatever it might have been, simply did not put him on a collision course with the Germans. As this chapter will show, Germans turned out to be Airy's strongest allies.

One could hardly guess, from the standard retellings, that there were actually several internationally influential German astronomers at the time, who had a decisive role in the Neptune story. The goal of this chapter is to explore their role, and, in particular, the ways in which they contributed to the discovery. As often is the case in science, a discovery might be attributed to a single person, or perhaps two as occurs here with Neptune, but it almost never happened alone and in isolation. In the case of Neptune, this was particularly true.

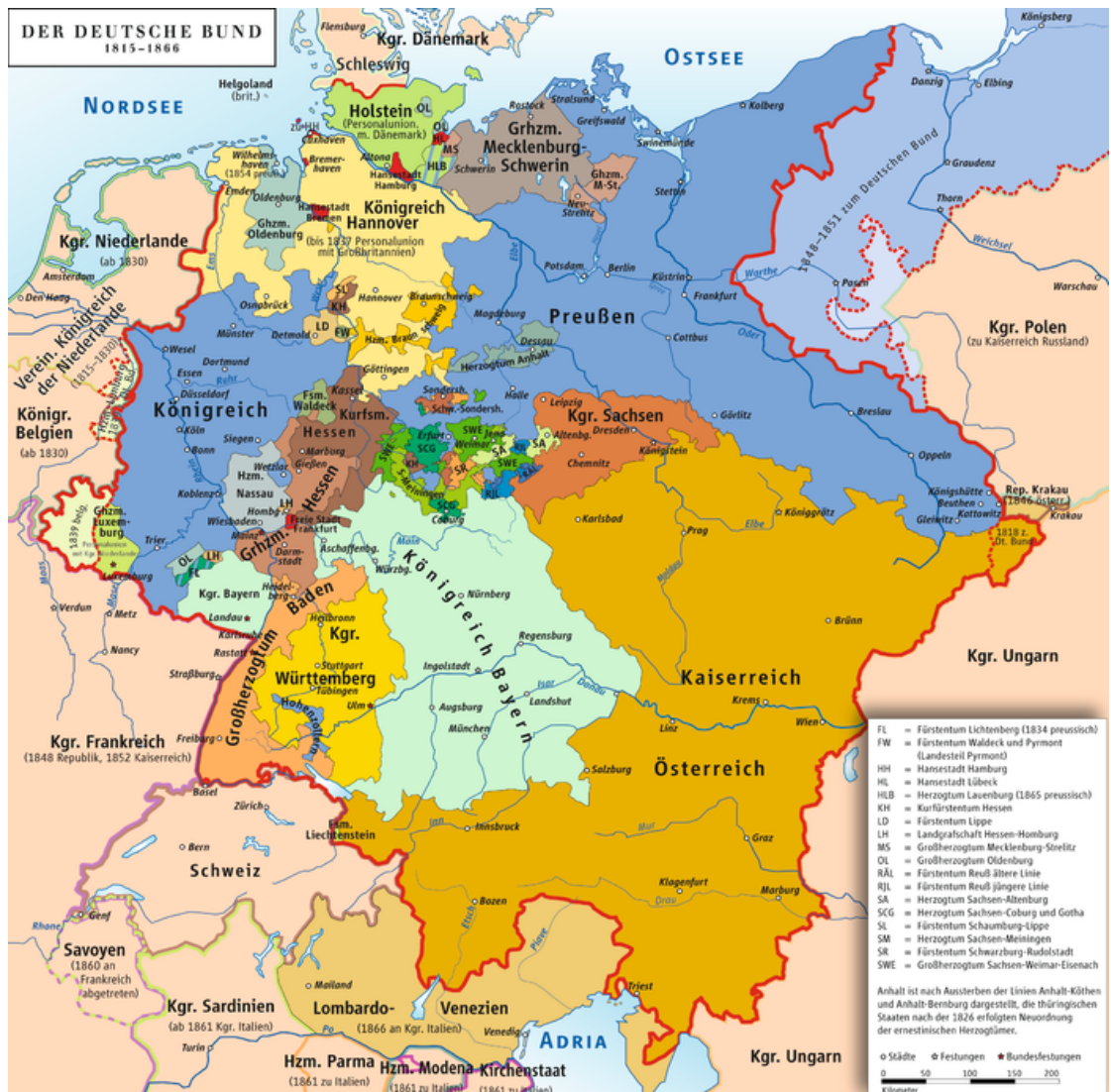

**Fig. 6.1** Map of German principalities in the first half of nineteenth century. The red line outlines the boundary of the German Confederation established at the Congress of Vienna in 1815, as an economic substitute for the Holy Roman Empire of the German Nation that was abolished by Napoleon in 1806. The development of the political situation after the discovery of Neptune slowly reshaped the confederation, first by creating a custom union without lands under the control of the Austrian Empire, then by addition of Schleswig, and finally with the unification of Germany under Prussian leadership in 1871 (Wikipedia commons. Credit: ziegelbrenne/ CC BY-SA 3.0).

It should be noted from the outset that by "German astronomy" we cannot refer to a monolithic bloc in the same way we can discuss the British or French astronomy. There was no "Germany", in the sense of a unified national state, in 1846. There was the "German Confederation" created by the Congress of Vienna in 1815 in the aftermath of the Napoleonic wars as a primarily economic substitute for the Holy Roman Empire of the German Nation, which had been dissolved in 1806. The German Confederation consisted of 39 independent states (Fig. 6.1), but did not include all of the German-speaking states and principalities in Europe. In addition to the arbitrary character of a political and economic entity created by fiat, the Confederation was weakened by the rivalry between the Kingdom of Prussia and the Austrian Empire.

Although Germans had continued to be divided by politics and religions of their principalities, they nevertheless enjoyed a unified language, a sense of national pride and heritage, and were steeped in a brilliant culture. It is in terms of the latter sense that we define "German astronomy," and designate as German astronomers all of those who belonged to this wider German cultural and linguistic circle. Some worked in observatories in the small and large German principalities across Central and Eastern Europe. Though some of those who contributed to the Neptune story were in what today is in modern Germany (Berlin), others were in German states that were powerful at the time, as most notably Prussia (Königsberg, Bresslau), but which are now in different countries and are known by different names (Kaliningrad, Wrocław). Some others were outside German principalities but now belong to modern Germany (Altona near Hamburg). Finally, there are those observatories, which, despite being led by German astronomers, were located then—and remain still—outside of German borders (Dorpat—now Tartu, Pulkovo).

As might be expected from their far-flung geographical range and many centres of astronomical activity, German astronomers had a correspondingly widespread influence. Owing to the splintered nature of the German political entities, princely lands, duchies, kingdoms, ecclesiastical territories and free imperial cities resisted centralization of academic efforts and enabled a growth of regional observatories. This is most likely the reason why German lands had more observatories than any other country in the world until 1870, when they were surpassed by the United States. In particular, during decade of Neptune's discovery as well as the rest of the nineteenth century, there were more German observatories than in France and Britain put together (Herrmann, 1973:51). If one wanted to search the skies for new planets, German observatories, or observatories led by German astronomers across Europe, offered a natural and a well-connected network, as has been discussed, for instance, by Chapman (2017:18–20).

**6.2 Laying the Groundwork**

As discussed in Chap. 2, the 18th-century discovery of Uranus was an unexpected—if not, in fact, a chance—event. William Herschel was systematically sweeping across fields of stars night after night in March 1781, seeking to make out which of them might be double stars. He noticed one that had a disk, and returning to it again, found that it had moved.

Nineteen years later, Giuseppe Piazzi applied exactly the same method and discovered Ceres. He was busy compiling his own star catalogue, but unlike Herschel, Piazzi, working together with his assistant Niccolò Cacciatore, was comparing his observations with stars in the catalogue by Nicolas-Louise de Lacaille, until he found one star that was not present in the catalogued stars of Taurus, and that also moved a bit across the sky the next night. Piazzi's discovery of Ceres differs from Herschel's discovery of Uranus, in at least two points. Firstly, while it was also serendipitous, it coincided with a targeted effort to look for planets elsewhere. Secondly, it became a showcase for the predictable power of the celestial mechanics and mathematical methods of the late 18th century.

Working at an observatory in the duchy of Saxe-Gotha-Altenburg, astronomer Franz Xaver von Zach was trying to assemble a network of observers, to search the sky for a hypothetical planet predicted to occupy the space between Mars and Jupiter. The prediction was based on the Titius-Bode distance sequence of known planets (see Chap. 5). In order to make an efficient search, von Zach realised that one needs good sky charts and tried to organize a network of astronomers who will share a burden of sky mapping. The agreement among von Zach's collaborators, Johann Hieronymus Schröter, Karl Ludwig Harding and Heinrich Wilhelm Matthias Olbers, was to chart stars down to the 9$^{th}$ magnitude over a region of the sky approximately covering 15º of latitude both south and north of the ecliptic. To do it efficiently a network of 24 astronomers was necessary so the work could be divided in zones of about 8º of longitude. In the end, although von Zach's "Celestial Police" turned out to be quite successful in discovering minor planets (they found three of the first four asteroids), Piazzi discovered the first minor planet while they were still trying to organize themselves.

The Celestial Police did not quite manage to deliver the planned all-encompassing charts of the zodiacal region. Harding created *Atlas novus coelestis* (Harding 1822) long after the group's efforts to search for possible planets had stopped. Harding's work stands in marked contrast to previous star charts, including the celebrated *Atlas coelestis* of John Flamsteed (Flamsteed 1729), *Atlas coelestis* by Johann Gabriel Doppelmayr (Doppelmayr 1742) and *Uranographia* produced by Johann Elert Bode (Bode, 1801). Whereas *Uranographia* has been referred to as "the pinnacle of gloriously engraved star-atlases" (Olson and Pasachoff, 2019:42), Harding's *Atlas novus coelestis* completely dispensed of pictorial extravagance. More to the point, it is historically significant as being the first sky chart published without drawings of the allegorical constellations, with an unprecedented purpose of being solely created in support of practical work. (Fig. 6.2).

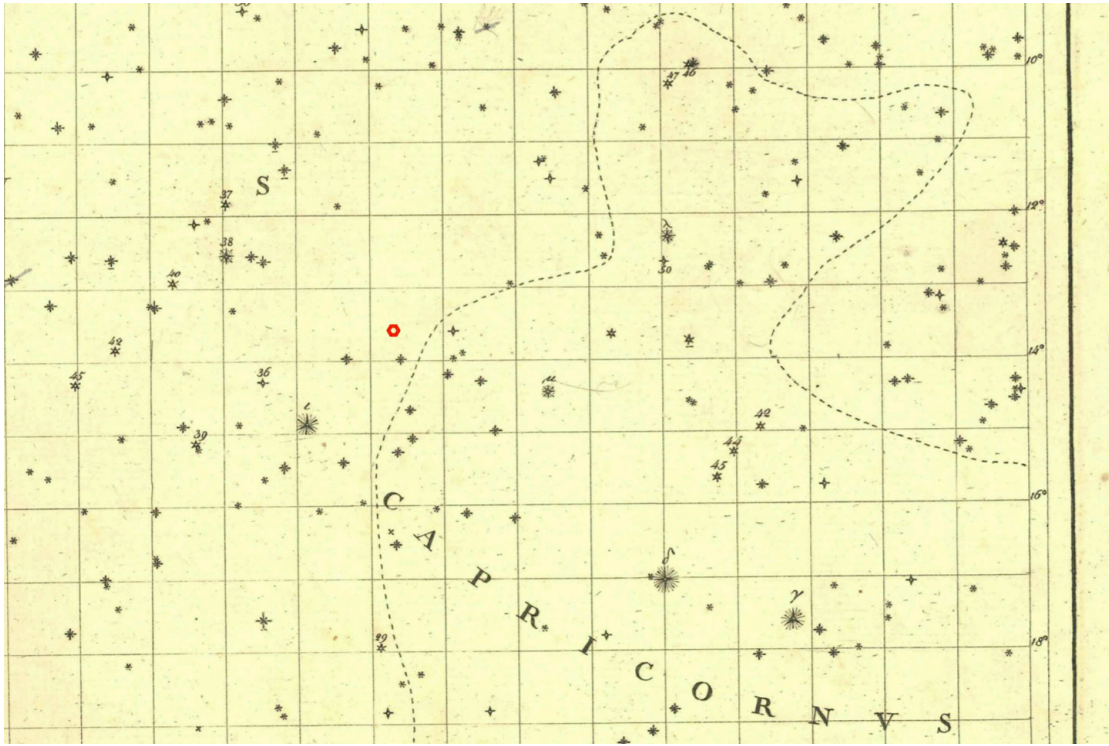

**Fig. 6.2** This star chart from the 1822 *Atlas novus coelestis* by Harding (1822) shows the region where Neptune was found. The red hexagon shows the location of Neptune at the discovery night. The incompleteness of this map can be judged by comparing it with Carl Bremiker's Hora XXI (Fig. 6.5). Note how constellations are marked only in a rudimentary way. (Credit: Library of Leibniz-Institut für Astrophysik Potsdam)

These events are not directly related to the discovery of Neptune some 45 years later. In retrospect, however, they indicate that the discovery of Neptune was a story in two parts. The first one was theoretical, belonging to what was then called "mathematical astronomy." The newly developed mathematical tools and, in particular, the confidence in the Newton's theory of gravitation ultimately emboldened the mathematical astronomers to put forward unprecedented predictions. The other part was empirical and consisted in preparing the ground, such as charting the skies to facilitate location of transiting bodies, and the actual observations necessary for any genuine discovery of astronomical objects. In both of these activities the most prominent German astronomer of the generation, Friedrich Wilhelm Bessel, took a central role.

### 6.3 Bessel's verification of Newton's theory of gravitation

Bessel was born in Minden, then part of several unconnected principalities belonging to Prussia, but in the vicinity of other major German political entities, Bremen and Hanover. Not a very good Latin scholar, but inclined to natural sciences, he asked his father to allow him to choose a business profession, one that would allow him to perform calculations (Schmeidler, 1984:11). In 1799, he was sent as an apprentice to Kulenkamp & Sons in Bremen, the trading company for which he worked for seven years. While the company appreciated his talents and even started paying him a decent salary, Bessel quickly recognised that the work in the office, even of an expanding business, could not occupy his mind fully. In order to understand the intricacies of the sea trade, especially determination of the location of a ship, he started studying the art of navigation, which led him to trigonometry and more complex mathematics, and finally to astronomy. Bessel was a self-taught prodigy who happened to be at a right time at the right place.

At the start of 1801 Piazzi discovered Ceres and during the same year Johann Carl Friedrich Gauss predicted its location, which helped Olbers and von Zach to find the asteroid. Then on 28 March 1802 Olbers discovered the second minor planet, Pallas. The Celestial Police was in full swing with Harding discovering Juno in 1804 and Vesta in 1807 by Olbers again. The discovery of Pallas was of high importance for Bessel; not only did it captivate his mind, it also put his interests firmly into domain of astronomy, but he also knew (at least on sight) its discoverer.

Heinrich Wilhelm Olbers was a well-known physician in Bremen with a practice below and an observatory on top of his house. In 1804, then 20-year-old Bessel gathered courage and approached Olbers with his calculation of the orbit of Halley's comet based on the 1607 data by Thomas Harriot (1560–1621) and Nathaniel Torporley (1564–1632), unearthed and published by von Zach in 1793 in *Astronomisches Jahrbuch* of the Berlin Observatory (Hamel, 1984a:16–17). Olbers was impressed by the work of young Bessel, Gauss as well, and while Bessel continued to work for Kulenkamp & Sons, it was only a question of time when he would change his profession.

Bessel first moved to a private observatory run by one of the key members of the Celestial Police, Johann Schröter in Lilienthal, rejecting an offer of a secure and a well-paid job at Kulenkamp. Bessel replaced Karl Harding in the position of an assistant. (Harding, following the fame he had achieved as the discoverer of Juno, had accepted a position as professor of practical astronomy at Göttingen.) Schröter's observatory was well equipped with telescopes of "exquisite beauty," as Bessel himself declared (Bessel, 1810:4). It also had one of the largest telescopes in the world at the time, with a mirror 48.4 cm in diameter, well suited for Schröter's passion of observing the surface of the Moon and planets (Hamel, 1984a:22). Nevertheless, in 1810, after 3 years with Schröter, and now aged 26, Bessel moved to Königsberg as the director of a newly founded observatory and became the professor at the Albertus University. Only then did his financial means start equating with what he could have earned as a businessman.[1]

Bessel's move from Lilienthal was timely; Schröter's Lilienthal observatory didn't survive the retreat of the French soldiers in 1813, who destroyed or carried off clocks, telescopes, and other astronomical instruments (see Chap. 2.4). In such wartimes, it is remarkable to consider that the government of Prussia, after the defeat by Napoleon in 1807 and effectively being occupied by the French, was thinking of building astronomical observatories. This effort should probably be viewed within the larger reforms within the Prussian state, in which Wilhelm von Humboldt (1767–1835) led an educational reform. Next to establishing the first university in Berlin (today's Humboldt University), he also convinced the King that the east Prussian capital Königsberg should have a well-equipped observatory led by a prominent astronomer. The choice of Bessel, by that time sought after for positions in Leipzig, Griefswald and Gotha (Schmeidler, 1984:17), was exceptionally longsighted.

Of the German astronomers connected to the Neptune story, Bessel is nowadays by far the best known. His contributions to mathematics, geodesy and astronomy are numerous. To a general scientific audience, he is best known for his estimate of the oblate shape of the Earth, his solutions to certain differential equations, his analyses of instrumental errors and their effects on observations, and his first measurement of stellar parallax (61 Bessel 1838), beating to it Wilhelm Struve and Thomas Henderson (1798–1844). He also predicted the existence of an "invisible" companions to stars Sirius and Procyon (Bessel 1844a,b).

---

[1] When Bessel was moving to Lilienthal, Kulenkamp & Sons offered him a salary of 700 Thalers per year, while for the appointment in Königsberg he got 800 Thalers per year together with a house and heating expenses covered (Schmeidler, 1984:18). This could be compared with the wages (Wulf, 2016:124) of a skilled worker (up to 200 Thalers per year), Alexander von Humboldt's pension by the King of Prussia (2,500 Thalers per year) and Wilhelm von Humboldt's salary as the ambassador in Rome (13,400 Thalers per year). When Johan Franz Encke became the director of the Berlin Observatory in 1825, he had a salary of 2,000 Thalers, of which 300 Thalers he earned as the secretary of the Berlin Royal Academy of Sciences, but his compensation did not include the accommodation (Bruhns, 1869:115).

It is less known, however, that Bessel spent a considerable effort on a theory of gravity directly motivated by the anomalies of the motion of Uranus. This is not surprising as nothing of Bessel's theory was published, and it only resurfaced in unpublished papers and letters Bessel exchanged with fellow astronomers including Olbers, Heinrich Christian Schumacher, Johann Franz Encke, Wilhelm Struve, and Gauss. It was one of those scientific projects that seem to lead nowhere, and in hindsight looks to be a completely incorrect path. For Bessel, however, this was a theory worthy of investigation, but once he was convinced that it didn't work, he completely abandoned it. He was convinced that the Newton's law of gravitation can be used at the vast distances of the space, including other stars, and that there is only one solution to the Uranus problem.

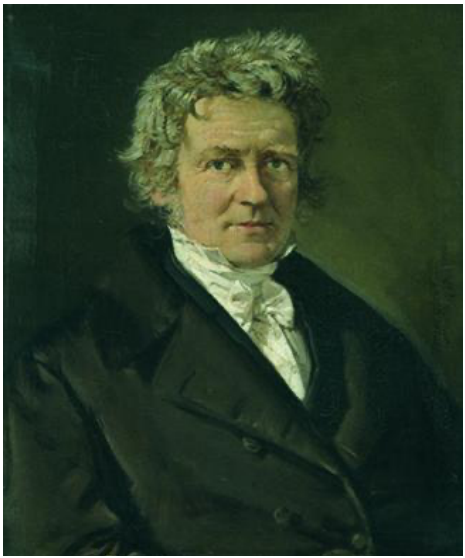

**Fig 6.3** Friedrich Wilhelm Bessel (22 July 1784 – 8 April 1846). Bessel was intrigued by the Uranus problem for 30 years, observing it and testing Newton's theory of gravity until he was convinced an external planet must be perturbing Uranus's orbit. This oil painting is from 1839 by Christian Albrecht Jensen, who also painted a number of other prominent German astronomers of the time, and is in the Ny Carlsberg Glyptotek Collection (Wikipedia, Public Domain).

Bessel's hypothesis for the deviations of Uranus from its predicted path was that Newton's law of gravitation should be modified. He asserted in an unpublished manuscript (Bessel 1823a) that "a planet can act upon two bodies not belonging to itself, at equal distances with unequal accelerating forces, or, according to previous use of language, with unequal masses."

Jürgen Hamel (Hamel, 1984b:280) presented this quotation in 1984 soon after the discovery of Bessel's unpublished manuscript. Bessel's idea was that the gravitational action of one body on another depended on the nature of each body. Thus, one can assume that the mass with which Saturn acts on Uranus is different from the mass with which Saturn acts with the Sun (or Jupiter). If this idea were true, then the anomalous motion of Uranus might be accounted for on the basis of this modification of Newton's law. In particular, this issue was recently raised by Alexis Bouvard (1767 – 1843) who showed that one could either satisfy the old (pre-discovery) or new (post-discovery) locations of Uranus with any given orbit, but not both at once (see Chap. 2.2 and 2.4; Bouvard 1821) Olbers, who seems to be the only one among Bessel's correspondents to show an interest in this theory, called it "*Wahl – Anziehung*" or "Selective Attraction" theory (Hamel, 1984b:279).

Bessel, however, did not use Bouvard's data. Instead he asked his students Friedrich Wilhelm Argelanger (1799–1875) and Otto August Rosenberger (1800–1890) to reduce the known observations of Uranus and derive the deviations, respectively. The data stopped with 1821, and the manuscript is not finished; Bessel did not seem to have been able to solve the problem of Uranus motions using the selective attraction hypothesis. Instead, he abandoned the project and moved onto other matters.

Bessel's abandonment of selective attraction is evident from his letters, notably on 2 October 1823 he wrote to Encke that he was not satisfied with his results. The derived masses for Saturn (as compared to the mass of the Sun) was 1/2482.2 when it interacted with the Sun, and 1/2427.7 when it interacted with Uranus (Hamel, 1984b:281). The later value was problematic as it was not compatible with the mass required to explain the interaction of Saturn with Jupiter, a problem already (mostly) solved in the eighteenth century (Wilson 1985). Soon after, on 9 October 1823 in a letter to Olbers, Bessel wrote (Hamel, 1984b:282): "My investigations into the motion of Uranus had to be concluded, because they have led to a result which makes further investigations on Jupiter and Saturn necessary."

Bessel's incursion into doubting the universality of the Newton's law was not confined only to the motion of Uranus. He performed numerous experiments testing the validity of the law of gravitation, experimenting with specially devised pendulum experiments. He investigated the effects of the increase of the moment of inertia of the pendulum due to air resistance, thus increasing the precision of the measurements. Inspired by the Uranus problem and his selective attraction hypothesis, following Newton's experiments, he also tested if by using different materials the oscillations of the pendulum would change. He used gold, silver, lead, iron, zinc, brass, marble, quartz, water and meteorite iron and stones (likely the first experiments using meteoritic materials), and concluded that the gravity does not depend on the type of material to within a fractional error of $6 \times 10^{-4}$ (Hamel, 1984a:48), an improvement of two orders of magnitude on Newton's result (Bessel, 1831, 1832).

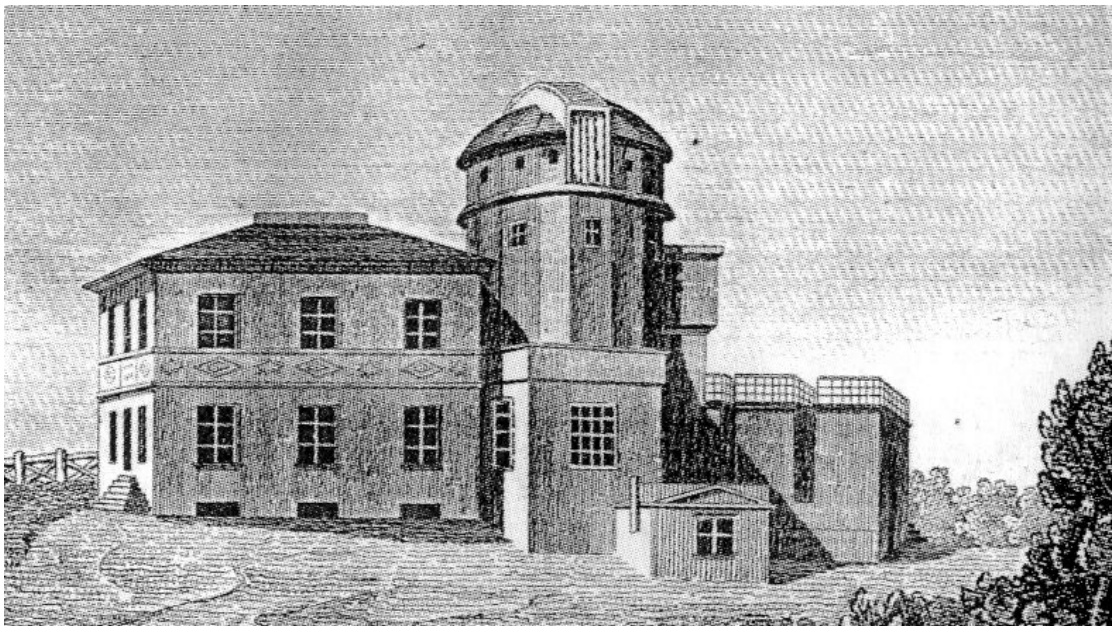

**Fig. 6.4** Königsberg Observatory in 1830. Bessel chose the location of the building and oversaw its construction. As the Prussian government delayed funds for the building, Bessel threatened to move to the observatory in Mannheim. The Prussian government backed down and provided necessary funds. Napoleon, passing with his armies through Königsberg to launch the attack on Russia apparently asked (Schmeidler, 1984: 20): "Does the King of Prussia still have time to think on such things?" (Wikipedia, Public Domain).

Although Bessel's experiments were performed several years after Bessel abandoned his work on the Uranus problem using the selective attraction hypothesis, that was not the end of Bessel's fascination with Uranus. As soon as he equipped his new observatory in Königsberg (Fig. 6.4) with decent instruments, Bessel started a systematic observing campaign of the planet. The first campaign was between 1814 and 1818, while the second extended a full 17 years from 1820 to 1837. At least the later part of the second campaign was fuelled by the conviction that Uranus motions could only be explained by an existence of an external planet. Already in 1824, in a letter to Gauss dated 14 June 1824, he mentioned this possibility, although sceptically, as an alternative to his failed selective attraction hypothesis (Hamel, 1984b:282): "I did not believe this; if one does not want to assume unknown, disturbing planets."

**6.4 Bessel's unfinished project**

It is not clear at what point in time Bessel convinced himself that it must be an external planet causing the perturbations. Peter Andreas Hansen, a Danish astronomer who spent most of his career as the director of the Seeberg Observatory in Gotha, claimed that he was always convinced on the existence of an external planet, contrary to Bessel (Wattenberg, 1959:3). Next to observing Uranus continuously, it seems Bessel actively came back to the topic in 1837 when he gave to an assistant Friedrich Wilhelm Flemming the task of re-reducing all available Uranus data, observed at Greenwich, Paris and Königsberg. Flemming's colossal task of reducing 361 observations (148 from Greenwich, 70 from Paris and 143 from Königsberg) and comparing them with Bouvard's tables was posthumously published by Schumacher (Schumacher 1850).

Bessel must have expected a big breakthrough from Flemming's work. In a public lecture on 28 February 1840 in Königsberg (Bessel, 1848:447), Bessel strongly expressed his opinion that an external planet will be found: "Additional explanations will aim to assign an orbit and mass for an unknown planet beyond Uranus, which may not be visible because of its exceeding faintness, which is such as to produce disturbances of Uranus, which cannot be put into agreement with existing observations."

He furthermore even mentioned Flemming's work in a postscript to a letter to Johann Gottlieb Galle dated 30 April 1840 (Wattenberg, 1959:2). The fact that Bessel himself wrote to Galle at all is remarkable: Galle at that time was just an assistant to Encke and the letter was written to congratulate Galle on discovery of three comets. A few days later, replying to a letter of Alexander von Humboldt who asked him about news on the "planet beyond Uranus," on 8 May 1840, Bessel was a bit more cautious, saying

> I think I have already told You that I have worked extensively on this; it has not progressed further than the certainty that the existing theory, or rather, its application to our knowledge of the Solar System as we know it, is not sufficient to solve the riddle of Uranus. However, in my opinion, therefore, one should not regard it as unsolvable. First, we need to know exactly and completely what is observed of Uranus. One of my young assistants, Flemming, has reduced and compared all observations; and with that I now have the facts in full. (Wattenberg, 1959:5)

After bringing Humboldt up to date with the issues with Uranus perturbations, Bessel continued

> I therefore thought that there would come a time when one would find the solution of the riddle: perhaps in a new planet, whose elements could be recognised from their effects on Uranus and confirmed by them on Saturn. I have been far from saying that the time has come: I will only try to see how far the existing facts can lead. It is this work that has accompanied me for many years, and for which I have pursued so many different views, that their end will excite me, and therefore will be brought about as soon as possible. I have great confidence in Flemming who will continue the same reduction of the observations he has now made for Uranus, for Saturn and Jupiter in Danzig, to which he is called. It is fortunate, in my opinion, that he (for now) has no means of doing observations and is not obliged to lecture. (Wattenberg, 1959:5)

One interesting difference between the Bessel-Flemming approach and those of Le Verrier and Adams is that Bessel was not satisfied with knowing the deviations of Uranus only. He was pushing Flemming to investigate those of Saturn and Jupiter as well. No doubt this would have made the calculations much more complex, and as Le Verrier showed, not really necessary. Wilhelm Flemming, however, died soon after his move to Danzig (Gdanks) in 1840 and the actual work stopped.

There are two further instances in which Bessel is connected to Neptune in his last years of life. In the aftermath of Neptune's discovery, John Herschel announced in *The Athenaeum* (Herschel, 1846:1019) his evening discussion with Bessel about the possibility of existence of the external planet, to which Bessel said that

> the motions of Uranus, as he had satisfied himself by careful examination of the recorded observations, could not be accounted for by the perturbations of the known planets; and that the deviations far exceeded any possible limits of error of observations.

Herschel concluded that Bessel did not yet try to actually locate the planet in the sky, but also mentioned that Bessel sent him a letter on 14 November 1842 saying: "Uranus is not forgotten."

The evidence is, however, that Bessel returned only once more to the question of Uranus and that was in a letter to Gauss written on 8 November 1843 (Bessel to Gauss, 8 November 1843; Breifwechsel 1880:564–569). At some point earlier Gauss asked Bessel for a suggestion of a topic that could be proposed as a prize question by the Royal Academy of Göttingen. Bessel's answer was "*Wenn Sie Uranus zur Preisaufgabe machen…*" (If you make Uranus the prize question). In the same letter Bessel went back to describe his failed attempt in 1820s on the selective attraction hypothesis and recognised that his investigations had stopped after Flemming's death. Bessel's conclusion was "*Ich halte den Uranus allerdings für einen sehr geeigneten Gegenstand; aber ich fürchte, daß dies viel Aufopferung fordern*" (I consider Uranus to be a very suitable object; but I fear that this will require much sacrifice).

Gauss took notice of Bessel's suggestion and in the 1845 edition of the *Proceedings of the Royal Academy of Göttingen* covering years 1842–1844, the Uranus problem was put forward as a prize topic (Abhandlungen, 1845:X). The deadline for the submission was November 1846. In some previous works (starting with von Lindenau, 1848:235) on the discovery of Neptune, the announcement of the prize is mentioned to be 1842, but this earlier date is incorrect. Bessel's letter is dated on 8 November 1843, and there is a letter dated 4 December 1843 in which Gauss informed Schumacher that he had the newest prize questions (Gauss to Schumacher, 4 December 1843; Briefwecshel, 1863a:195). The publication of the prize in early 1845 is also consistent with the fact that Adams became aware of it (Adams, 1846:149), most likely before the Long Vacation in the summer 1845 when he produced the Hyp I solution.

Of all German astronomers Bessel worked the most on the problem of Uranus. His interest in the problem spanned most of his career, from 1814 when he started observing the planet, to at least 1843 and the letter to Gauss suggesting it as a prize topic. He approached the problem with an open mind and used it as a test of the theory of gravity. When that did not work, Bessel accepted that there might be yet another planet beyond Uranus. Being an observer par excellence equipped with state-of-the-art instruments, he performed his own observations and set students to the task of analysing this data.

Bessel was not the first one to speculate about an external planet, but his approach was meticulous. Perhaps even too much so, as adding the deviations of Saturn's motions would have made the problem even more complex. Flemming's death in 1840 was certainly a big setback for the work, but 1840 was an exceptionally hard year for Bessel. He lost his old friend and father figure of Olbers who died in March; then in October he received the news of the death of his son Wilhelm in Berlin. Even if picking up Flemming's work in the last years of life might have been too much for Bessel, his work on charting the skies left a legacy that eventually resulted in the discovery of Neptune.

### 6.5 Charting the skies

Regardless of the way the story of the discovery of Neptune is told, its central point is that an accurate star chart was necessary to find the planet in a timely manner. More detailed accounts like to dwell also on the chart's presence in Berlin and its absence in Cambridge. The map in question is the *Hora XXI* (hour 21 in right ascension) of the *Akademischen Sternkarten* made by Carl Bremiker (1804–1877) for the Royal Academy of Sciences in Berlin (*Königlichen Akademie der Wissenschaften zu Berlin*). Figure 6.5 shows the map kept at the library of the Leibniz-Insitut für Astrophysik Potsdam, the successor of the Berlin Observatory.

The story of the making of that map, as well as the whole series of star charts, started in 1824 with a letter from Bessel to Johann Elert Bode, then the director of the Berlin Observatory. In a letter dated 18 August 1824, Bessel asked Bode if he thought the Berlin Academy (of which Bessel was a member) would be willing to support a project of charting the skies. Bessel's point was that only with maps of all stars down to the 9th or 10th magnitude, in the ecliptic region, would it be possible to look for additional planets.

Bode answered enthusiastically on 21 September. By 15 October Bessel dispatched his proposal letter to Berlin Academy (Bessel's letter to the Academy is reproduced in the catalogue that accompanied the star charts; Cataloge, 1859:III–V). The Academy enthusiastically accepted Bessel's proposal during a meeting on 4 November 1824 (Hamel, 1989:12). The project was officially launched in 1825, and 34 years elapsed before it was successfully concluded.

**6.5.1 Bessel's star chart idea**

The purpose of the maps for Bessel was straightforward: to help determine precise positions of comets or planets, and to discover all major planets ("Hauptplaneten") of the Solar System. Even though no new minor planets had been discovered since Olbers spotted Vesta in 1807, Bessel's experience with star catalogues, which listed stars that were no longer visible, gave him hope that more planets could be found. (Cataloge, 1859:III).

Bessel's idea of star charts as tools for discovery of comets and planets was not new. He acknowledged in the proposal that the Lilienthaler Astronomische Gesellschaft (part of the Celestial Police centered around Schröter's observatory), as well as the Astronomical Society of London (the forerunner of the Royal Astronomical Society), had map-making as their mission. The problem was that the existing catalogues or sky atlases, such as those of Flamsteed, Bradley, Piazzi, Lalande, Mayer, and Bode were incomplete both in spatial coverage and brightness. They had regions that were under- and over-crowded with stars, and simply did not extend to stars beyond the 8th or 9th magnitude that were deemed necessary (Bessel 1823b). Furthermore, the actual maps were often more works of art than useful tools for examining the skies.

Bessel's main idea was essentially what von Zach and the Celestial Police were supposed to make: a set of charts covering the ecliptic and divided into zones one hour (in right ascension) wide, mapping positions of all fixed stars brighter than a specified magnitude. In 1822, Harding published the only existing set of maps of a similar kind, but those maps, which served as a template for Bessel's charts, were still incomplete (compare Figs. 6.2 and 6.5 around the location of Neptune on the discovery night).

Bessel's proposal to the Academy ended with five propositions (Cataloge, 1859:IV–V):
1) The maps should include stars in the zone 15° north and 15° south of the ecliptic, divided in 24 zones of one hour width, including 4 arcminutes (1°) overlaps on both side of the hour. Each map should be divided in a grid of 510 squares, each of 1° width and height.
2) As a reward for making the map, each astronomer should receive a prize (as a medal) worth 20 to 25 Dutch Ducats.
3) The Academy should set up a commission to take charge of the project and the awarding of the prizes.
4) Each map would be cast in copper so that it could be reproduced at will at a later date.
5) A summary of the project [a flyer] would be made available to the public.

Bessel understood creating accurate star charts was complex and not very scientifically exciting work suitable to be done in parallel by a number of "Freunde der Wissenshaft" (friends of the science): "This is a business which requires much time, but few resources and even less knowledge, and which is therefore most suitable for lovers of astronomy and for those astronomers whose observatories are not so fully equipped, that every night deprived of science will not cost it a considerable sacrifice."

His proposal was quite clear on who should be employed to make these charts:

> Considering the number of young astronomers, or older not fully equipped, which is currently available, one must also wish from this side that serious means are taken now to leave such a large and glorious monument for posterity. (Cataloge, 1859:V)

## 6. "That star is not on the map": The German Side of the Discovery

Staubermann (2006) describes the process of remaking one of the Academy charts using Bessel's suggestions and instrumentation available in the first half of the 19th century. His conclusion was: "The first ten or so stars posed a challenge, the first fifty called for skills that needed aptitude, such as the ability to copy the stars from the telescope on paper, but after about eight stars the exercise became dull and boring."

Indeed, among the astronomers who volunteered at the beginning there were already famous names, such as Wilhelm Struve and Alexis Bouvard, but many of the maps were made by Bessel's "friends of the science", such as Carl Bremiker, Thomas John Hussey, Karl Ludwig Hencke (1793–1866), observers whom one would nowadays most likely call serious amateur astronomers.

Bessel thought out the full process of chart making, and in addition to his initial proposal to the Academy he also provided details of how to observe in practice. The framework Bessel set up, next to the detailed description of what each map should look like, consisted also of how astronomers should make their observations and put them onto the map. The grid of 510 squares of 4 minutes of time (1º) width and 1º height should first be filled with stars observed in Königsberg (by Bessel), Paris (Lalande 1801), and Palermo (Piazzi 1814), adjusted to the epoch of 1800. The observations would then commence: each observer would use a standard comet seeker (Bessel suggested a Fraunhofer refractor of 750mm focal length or a Dollond of 400mm focal length) together with a cross-wire micrometre. Such a telescope would provide a field of view comparable to (or larger than) one of the squares of the grid and allow for mapping stars of 9th or 10th magnitude.

The observer's task would be to "by eye" ("blosen Augen") plot fainter stars (those that are not in the catalogues) onto the map. For each fainter star (dimmer than 7th magnitude) there should be an indication if it was observed once or twice. Special care should be taken depicting binary stars, and charts should be verified by comparison with heavens (Cataloge, 1859:IV).

Bessel himself observed 32,000 stars down to the 9th magnitude in the same -15º to +15º zone, but he realised that more was needed to improve on previous star charts. The main reason was that three of the four discovered asteroids were fainter than the 8th magnitude so, therefore, 9th or 10th magnitude was a necessary limit. Going much deeper would increase the number of stars considerably, as James Challis experienced in his search for Neptune, and render the project unfeasible. Nevertheless, even down to just 8th magnitude, the number of stars was still large, about 2,000 to 3,000 per zone. It was imperative to find a working compromise between the precision (both in terms of the recorded spatial location and brightness of stars) and the time required to map a zone.

Already in 1823 Bessel introduced a term "*Zeitersparnis*" (Bessel, 1823b:261). This "time saving" idea was his attempt to minimize the time needed for observing stars. What is striking for Bessel, the leader of precision astronomy and error analysis, is that he decided to sacrifice the precision of the measurement to reduce the time needed to make it.

At that time, to determine the precise location of a star, astronomers used a meridian circle to observe a star as it crossed the local meridian and at the same moment note precisely the local time. Meridian observations demanded expensive instruments (including good clocks) as well as an assistant to help with measuring the time. In such a way, Bessel, with the help of his student and assistant Argelander, was able to measure the precise position of three stars in a minute (Bessel, 1823b:262). Therefore, for map making, Bessel's advice was different, tailored for less equipped observations by a single astronomer. Entering fainter stars "by eye" onto a grid of pre-existing stars sped up the process tremendously but required multiple re-observations.

Bessel's student Carl Steinheil (1801-1870) designed an apparatus to help observers in plotting stars by eye (Steinheil 1826). It consisted of a wooden drawing board (see Staubermann, 2006:27 for a replica and its usage) with two brass rulers, set up perpendicular to each other and extending the full length of the board. The first ruler covered from -16 to +19 degrees (of declination) with intervals of 15', while the second covered 68' of right ascension in intervals of 15". With such a mapping system, it was possible to position stars with a precision of 1 arc minute in declination and 1 second of time in right ascension. Such precision was much less compared to that at which the stars in Bessel's, Lalande's, or Piazzi's catalogues were given, but it is the limiting accuracy at which a human eye can effectively position the star on a map of such specifications (Staubermann, 2006:28).

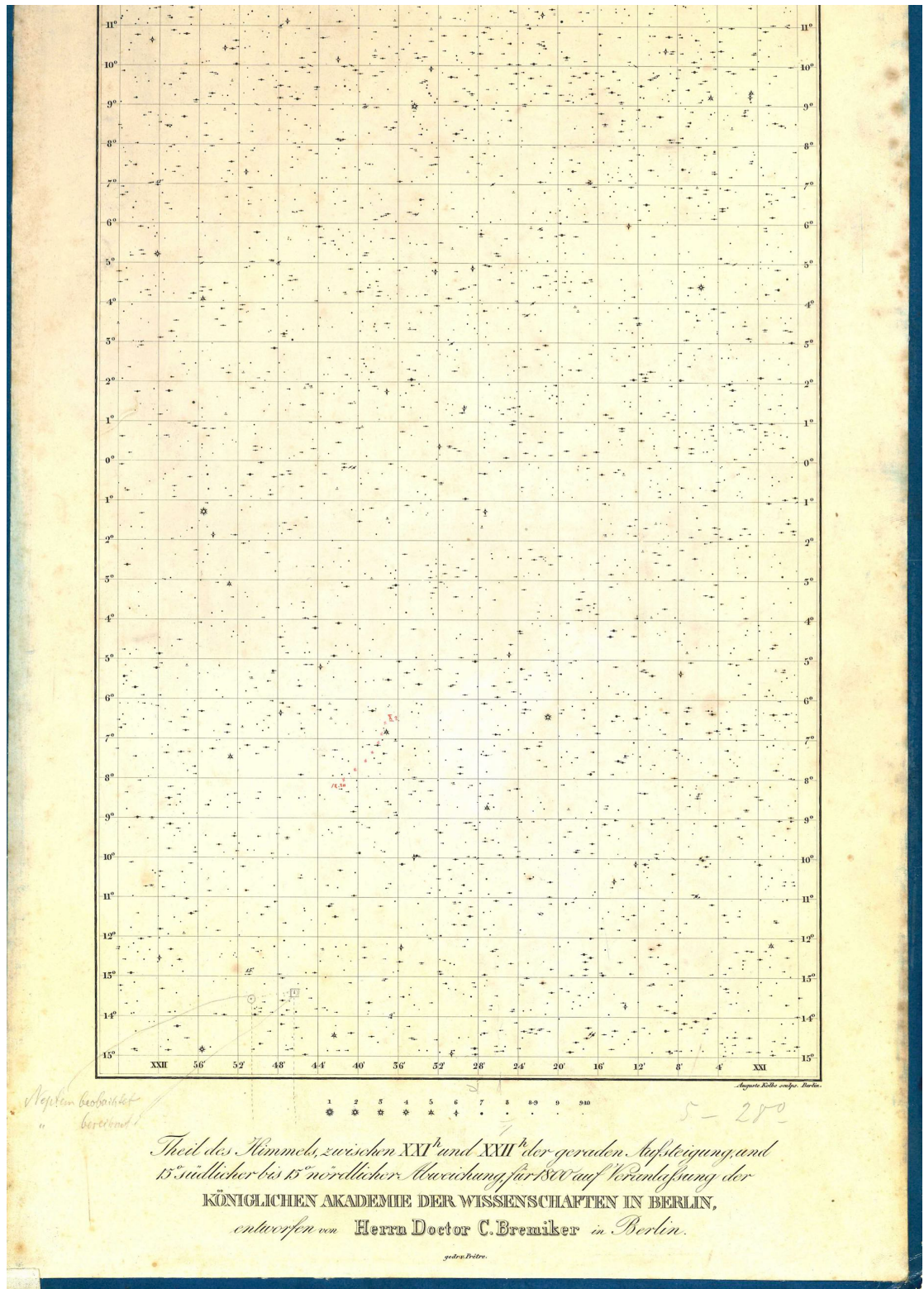

**Fig. 6.5** Hora XXI of the Royal Berlin Academy of Sciences assembled by Carl Bremiker, kept in the Library of Leibniz-Institut für Astrophysik Potsdam. Only part of the map with declinations between 11° and -15° is shown. Horizontal lines next to stars (two or one only) indicate the number of times each star was observed. The handwritten notes show the predicted (square) and observed (circle) locations of Neptune. The notes are attributed to Galle but were added some time after the discovery. The map continued to be used for regular work, as the red sequence of dots testifies. The main advantage of this map compared to Harding's (Fig. 6.2) is that it was complete to stars of the 9th magnitude, which is particularly noticeable in the vicinity of Neptune's location on the discovery date (Credit: Library of Leibniz-Institut für Astrophysik Potsdam).

In addition to plotting the relative positions of new stars, the observer had to record their brightnesses. Bessel's idea was to divide them into one magnitude bins down to the 8th magnitude and then to note stars of 8th–9th, 9$^{th}$, and 9th–10th magnitudes, each with a different symbol (see bottom of the map on Fig. 6.5). Even though it is possible for a trained observer to make a finer grading of stellar brightness, in fractions of magnitudes, that was not the aim for the charts. More important was to show how many times a fainter star (that have been located "by eye") had been observed.

**6.5.2 Star charts of the Royal Academy of Berlin**

Bessel had the foresight to see how complicated and costly[2] such a project would be and therefore asked for the support of the Royal Academy of Berlin. In hindsight he was completely right: the project lasted 34 years (counting from the proposal announcement). Twenty astronomers produced 24 maps in observatories across German lands (Kingdom of Prussia including parts of Poland, Kingdom of Bavaria, Kingdom of Hanover), as well as Denmark, Imperial Russia (Finland and Ukraine), the Habsburg Empire (Bohemia), the Grand Duchy of Tuscany, the Kingdom of Two Sicilies and the British Isles.

Fewer than one third of the maps were started and completed by the same person. For the majority of maps, two, three, or even four astronomers worked on them at some point. The total number of people involved was 36. The overseeing committee established by the Academy did not fare much better. All but one of its members changed during the three-plus decades of the project: a total of seven members served on the committee, of whom four died and one moved to a post in a different principality. That constant member was Johann Franz Encke (Fig. 6.6), who was elected in the commission (June 1825), as its chair, even before his arrival to Berlin in October 1825 (Hamel, 1989:12).

Encke was a veteran of the Napoleonic wars, in which he served as a Prussian artillery officer, twice stopping his studies in Göttingen under Gauss to join the war activities. After the war he worked at the Seeberg Observatory and became internationally famous by showing that comets discovered by Pierre Méchain in 1786, Caroline Herschel in 1795 and Jean-Louis Pons in 1805, and again in 1818, were the same object, and therefore establishing the existence of the second known periodic comet (the first being the Halley's comet). Another important point about Encke's comet, as it has become known since, is that, is that it has the shortest period (3.3 years) of any comet and its aphelion falls within Jupiter's orbit (Bruhns, 1869:59–65). The accuracy of Encke's calculation was tested in 1822, the year he predicted the comet would reappear. Indeed, Carl Ludwig Christian Rümker (1788–1862) spotted it from the Parramatta Observatory in Australia on 1 June 1822 (Bruhns, 1869:64).

The influence of Encke's discovery on the astronomical community of the time is best substantiated by the action of the Astronomical Society of London (which later became the Royal Astronomical Society), which awarded Encke its very first Gold Medal (Rümker was given the silver medal) in 1824. The recognition of the Royal Astronomical Society likely had an effect when Encke was considered for the directorship of the Berlin Observatory, on Bessel's recommendation (Bessel himself turned down the job as he was satisfied with his position in Königberg; Bruhns 1869:102). Encke was given the post of the director of the Berlin Observatory (upon retirement of Bode), professorship at the University and a position in the Royal academy. Even though he arrived after the beginning of the star chart project, he was crucial for its organisation and execution.

The project was officially announced in No. 88 of the *Astronomische Nachrichten*. The issue (Schumacher, 1826), printed and distributed in December 1825, contains Bessel's proposal to the Academy together with an example map, also prepared by Bessel. Bessel's name does not appear anywhere; the reader only knows of Bessel's involvement from a letter sent by Encke to Schumacher on behalf of the Royal Academy of Berlin. Only later, in No. 93, printed in April 1826, readers learned that the star chart effort was Bessel's project, as he provided additional instructions on how to make a map (Bessel 1826), followed by a description of the Steinheil's drawing board apparatus for efficiently plotting stars (Steinheil 1826).

---

[2] Estimated expenses were 3,000 Thalers, of which 2,000 Thalers would go as prizes to astronomers (Hamel 1989).

Encke's role in the completion of the project cannot be understated. His team of observers was spread across Europe. The team members, most of whom volunteered and some of whom were directors of local observatories, were often slow in answering his letters, and after years of non-responding would announce their withdrawal from the project. As the years progressed, Encke started relying on his students Carl Bremiker, Jakob Phillipp Wolfers (1803–1878) and Heinrich Louis d'Arrest to complete maps, some of them started years before.

The celebrated Hora XXI map was itself such a case (Fig. 6.5). Even before the official announcement of the project in the *Astronomische Nachrichten*, some volunteers were selecting their preferred hour zone. One of them was Otto August Rosenberger, a former student of Bessel and now a professor in Halle (Saale). He started working on the Hora XXI map in 1826 (see Hamel 1989 for a detailed history of Hora XXI). In 1828, Rosenberger informed Encke that the progress was slow due to weather conditions. In 1831, Encke sent letters to all astronomers still working on maps (only four maps had been completed by then), and Rosenberger replied that he was hoping to keep the deadline of 1833 suggested by Encke. By 1839 there were still no data from Rosenberger. After 14 years of waiting, Encke decided to change the astronomer in charge. He gave this work to Carl Bremiker, his former student who worked as an inspector for the Prussian Minister of Commerce. Bremiker actually took over the work on three maps for which he was supposed to receive 300 Thalers (somewhat more than envisioned by Bessel's proposal).

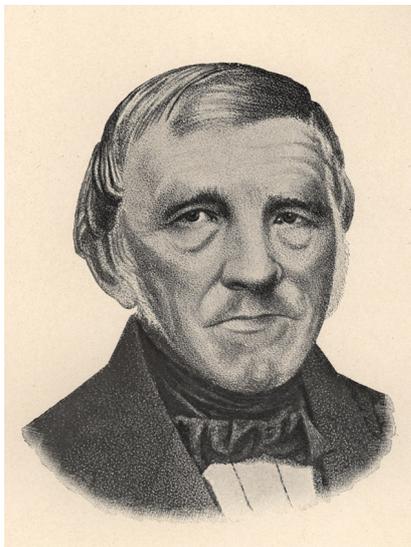

**Fig. 6.6** Johan Franz Encke (23 September 1791 – 26 August 1865). A veteran of Napoleonic wars, Encke became famous internationally for proving that several sightings of comets were consistent with belonging to a single short period comet, the first of its kind to be discovered. Observations of the comet's predicted return in 1822 from the newly established Parramatta observatory in Australia earned Encke the very first Gold Medal of the Astronomical Society of London (the forerunner of Royal Astronomical Society). Encke was awarded the second Gold Medal in 1830 for his work on the ephemerides at the Berlin Observatory, becoming the first recipient of more than one Gold Medals. Credit: Bruhns (1869) (Credit: The picture is based on a photograph and is the frontispiece in the 1869 biography of Encke (Bruhns, 1869)).

Rosenberger provided all the data that he had collected so far and Bremiker was able to complete the work within a year (from 1 January to 4 December; Hamel, 1989:17–18). It took another four years for the map to be checked and accepted by the Academy, and finally it was engraved in copper and printed on 9 November 1845. On 10 November, all the copies were ready to be distributed, but the map seemed to have faded from the memory of all, until resurrected by Heinrich Louis d'Arrest on 23 September 1846. To understand this lapse of time, it is necessary to understand that Hora XXI was just one map in a very large and, by then, a very long project.

Bessel, Encke, and the commission for the maps originally thought that it would take just a few years to complete the project. The plan was to print out a few hundred copies of each map and send them out in pairs (two different maps). There were several predefined recipients, both institutions and individuals. For example, the maps were sent to the Royal Society, the Royal Astronomical Society, Bureau des longitudes, Institut de France, Imperial Academy of Petersburg, Royal Academy in Naples, as well as directly to Gauss, Olbers and Schumacher. Until 1841, maps were distributed in this way, two at a time (Cataloge, 1859:IX).

When Bremiker finished Hora XXI, Encke was also expecting Hora XI from Palm Heinrich Ludwig von Boguslawski (1789 – 1851) in Breslau (Wrocław). Indeed, Boguslawski sent it on 12 June 1845. The first printed proof of Hora XI was sent back to Boguslawki in February 1846 for his examination, but he did not return it until 1851, as he was improving it (Galle, 1882a:220). It was not unusual that maps shipped together were finished at different times, but for the first 13 the usual time lag was of the order of a year or two. In the summer of 1846, even though time was slowly slipping away, there was no indication that a successful and timely distribution of Hora XXI and XI would not happen. To a large extent, the events surrounding the discovery of Neptune took over and only in hindsight one could ask the question: Why was Hora XXI not distributed earlier? From Encke's project management point of view, there was nothing special in Hora XXI, compared to any of the other 10 yet unfinished sky charts in 1846.

An interesting parallel can be made with Hora IV, the first of the Berlin maps that made history. On 8 December 1845, a retired postmaster and amateur astronomer, Karl Ludwig Hencke, discovered the asteroid Astrea. He did this exactly in the way Bessel thought the discovery of planets should happen; following observations with a good star chart. The map Hencke used, Hora IV, was made by Karl Friedrich Knorre (1801–1883) in Nikolayev (Mykolaiv, Ukraine). Hora IV was finished in 1834, but was sent out only in 1837. It is not clear what caused the delay in distribution (Cataloge, 1859:X), although it was likely induced by corrections and proofs. The map sat for eight years in drawers of learned societies and observatories across Europe until Hencke used it to discover the first new minor planet since 1807. Even the example of Hora IV did not suggest that Hora XXI was in any way special. Perhaps this is where the irony of the Neptune discovery lies: in addition to the stunning mathematical prediction, there was a lot of luck involved.

**6.6 The Discovery in Berlin**

Neptune was discovered at the Berlin Observatory in the night of 23 September 1846, based on the prediction by Urbain Jean Joseph Le Verrrier. It was found to be $1° 03' 06.7"$ away from Le Verrier's predicted position. (Gapaillard, 2015:55)[3]. With hindsight, it was a lucky find that surprised everyone involved—the observers, the observatory director, and the whole world.

There are many scenarios in which Neptune could have been discovered somewhere else at some other time. After all, it was observed several times before 23 September 1846. Galileo Galilei had observed it in 1613, Jerome Lalande published the observations of his nephew Michel Lalande (1766–1839) in 1795 in Paris, and John Lambert observed it as late as on 11 September 1846 in Munich (see Chapter 9.1). However, a series of opportune decisions led straight to the discovery in Berlin. The crucial decision started two decades earlier, with Alexander von Humboldt and Josef von Fraunhofer.

**6.6.1 Humboldt's masterpiece: a new telescope, a new building, and a paid assistant**

Josef von Fraunhofer (1787–1826), a famous physicist, is known as the father of spectroscopy. He developed the first modern version of this new instrument, the spectroscope, and he was also an optical manufacturer who created some of the best telescopes of his time. Bessel used a Fraunhofer heliometer in Königsberg to determine the parallax of 61 Cygni, just beating Wilhelm von Struve at the Dorpat (Tartu) Observatory who also used a Fraunhofer refractor (which was then largest in the world, having a lens 24 cm aperture and focal length 432 cm).

Fraunhofer made another refractor, identical to the one sold to Struve (who was financed by the Tsar of Russia). After Fraunhofer's death in 1826, that second telescope was offered for sale by Joseph von Utzschneider (1763–1840), the co-owner of the company. Gauss was interested, and he asked Schumacher, then visiting Munich, to find out more about the possibility of purchasing it.

---

[3] See the same reference for a discussion why there was a persistent statement in the literature that Neptune was found "within one degree" when the discrepancy is actually larger than a degree.

## 6. "That star is not on the map": The German Side of the Discovery

Schumacher's news was not very satisfactory for Gauss: the telescope was being offered to Joseph von Littrow (1781–1840), at the Vienna Observatory, possibly just for bargaining reasons (Knobloch, 2013:49–50), for a price of about 20,000 Thalers (Bruhns, 1869:179).

As it became clear that Gauss had decided not to purchase the expensive instrument, Alexander von Humboldt entered the bidding game. His approach was much more subtle, but he also asked Schumacher for help. Schumacher was supposed to convey to Utzschneider that Humboldt would like to buy the instrument and was willing to pay 8500 Thalers. What Humboldt did not want Utzschneider to know was that the money would come from the King of Prussia, with potentially a much "deeper" purse. In the end, Utzschneider accepted the offer, and Schumacher, while unable to help Gauss, played a crucial role in Humboldt's grand scheme to improve the Berlin Observatory. (Knobloch 2013).

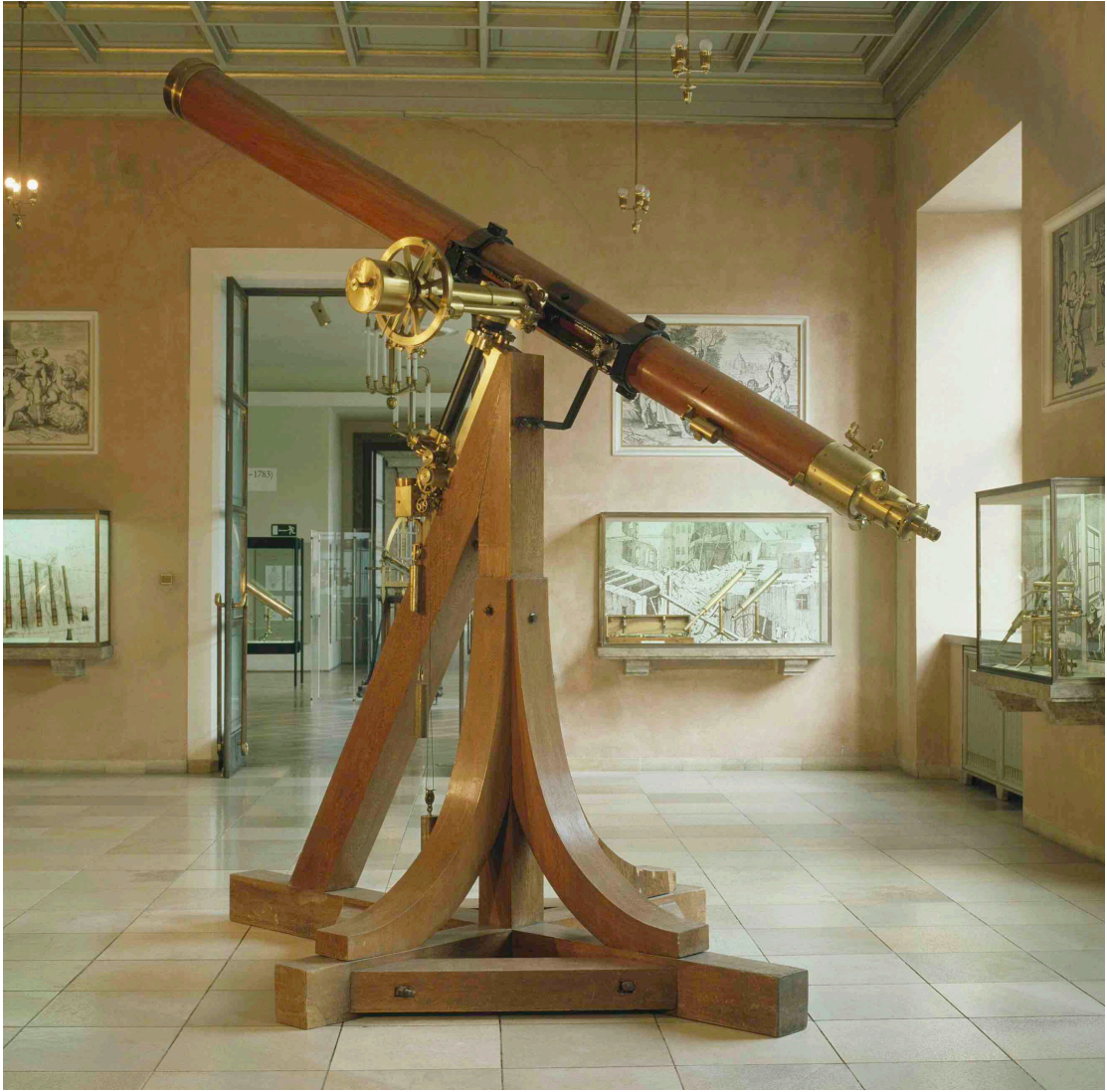

**Fig. 6.7** Fraunhofer telescope of the Berlin Observatory (identical to the one at the Dorpat Observatory) was used for the discovery of Neptune by Galle and d'Arrest. The refractor has a lens of 24 cm aperture and focal length of 432 cm. The telescope is kept in the Deutsches Museum in Munich. (Credit: Deutsches Museum)

Appointing a new director, then as now, was a good opportunity to renew instruments, refurbish buildings, and in general argue that more funding was necessary to keep up the good work and the prestige of the observatory. The origins of the Berlin observatory go back to the calendar reform of the Prussian state in 1699 (then still a dukedom) and the foundation of a Kurfürstlich-Brandenburgischen Societät der Wissenschaften (a forerunner of the Royal Academy of Berlin of Sciences) proposed by Gottfried Wilhelm Leibniz. The observatory was awarded the calendar patent, which provided the income for the observatory and the academy for most of the next century. The first building was constructed in 1711 in the Dorotheenstraße in the academy building tower above a former stable in the central part of the city.

Humboldt's idea was to obtain a first-class telescope and then to ask for a new building suitable for the telescope. Encke, who by 1826, started living opposite the old observatory (as Bode was still alive and occupying the director's apartments) was not too enthusiastic (Bruhns, 1869:179), but the opportunity arrived with the Fraunhofer telescope (Fig. 6.7).

In March 1828 the new refractor reached Berlin stored in the boxes that remained its home for several more years. According to Humboldt, such an expensive and excellent instrument needed a proper observatory building, and he turned to the king again. A new building was approved and Encke selected a location, which was part of a larger park without nearby houses. Already by mid 1840s, however, houses appeared around the new observatory building; even though their height was restricted, parts of the horizon visible from the observatory dome were obstructed (Bruhns, 1869:185–186).

The location of the observatory in present Berlin was at the crossing of the Enckestraße and Besselstraße, southeast of and not far from the Checkpoint Charlie. Today nothing remains of the elegant neo-classical building designed by Karl Friedrich Schinkel (Fig. 6.8). The building of the observatory started in October 1832 and the main house and the dome were finished by December 1835 (Bruhns, 1869:186–187). Encke moved in on 24 April 1835 and set up his older instruments just in time to observe the return of the Halley's comet. The new Fraunhofer refractor had to wait a bit longer in boxes: the unpacking started on 28 September 1835 and five days later it saw "first light". Encke's report about the excellent quality of the new telescope (and the new observatory building) appeared in February 1836 in *Astronomische Nachrichten* (Encke 1836).

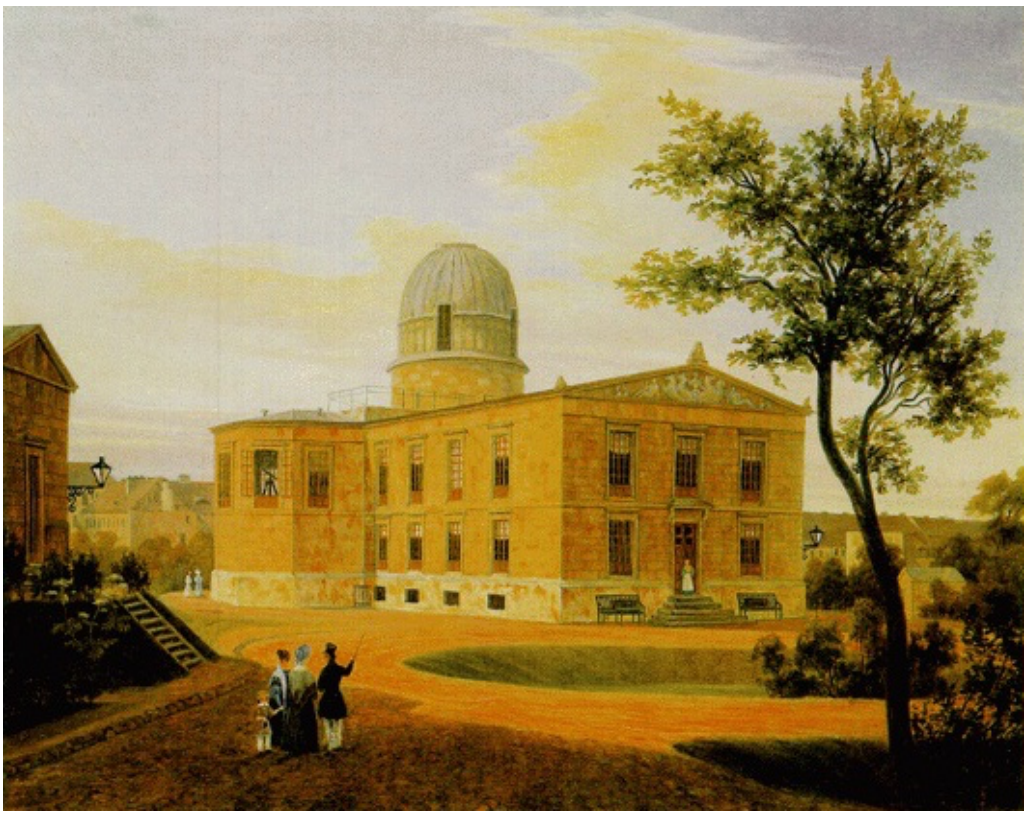

**Fig. 6.8** The new building of the Berlin observatory by Carl Daniel Freydanck (oil on canvas, 1838). The architect of the observatory was Karl Friedrich Schinkel who also designed several other buildings in Berlin. The observatory was opened in 1835. The main drive to build the new observatory was to provide an appropriate place (the dome) for the new Fraunhofer telescope (Wikipedia Public Domain).

Humboldt's deal of getting a telescope and then building the new observatory was extended to include a paid assistant for the director. (Wattenberg, 1962:34). The paid assistant, who also lived at the observatory, moved in just three days after Encke (Wattenberg, 1963:34). His name was Johann Gottfried Galle (Fig. 6.9).

Galle was born in Pabsthaus, next to the village of Radis in the vicinity of Wittenberg in 1812. His family's business was producing tar, but there seemed to be a clear indication from an early age that Johann Galle should not continue his father business but go to school instead. He finished the gymnasium in Wittenberg in 1830 at the top of his class and matriculated at the Berlin University as a student of mathematics and philosophy. Encke was Galle's teacher and, in later study years, Galle started going to the old observatory and contributing calculations for Encke's *Berliner Astronomisches Jahrbuch* (Wattenberg, 1963:26). His relations with Encke must have been very good, as in winter of 1834 Galle wrote to Encke asking if there were a chance for a paid assistantship in the new observatory that was being built (Wattenberg, 1963:32–33). The positive answer was sufficient for Galle to resign as a teacher in Friedrichs-Wederschen Gymnasium in Berlin, take the post of the assistant and devote his life to astronomy.

Over the next decade, the newly installed Fraunhofer refractor proved to be an exceptional instrument. Using the telescope, Encke in 1837 discovered a gap in the A ring of Saturn, while Galle in 1838 discovered the darker C-ring. Galle also found three comets between December 1839 and March 1840, which attracted the attention of the wider public and professional peers. First Galle received the Gold Medal from the Prussian king (14 March 1840). Then, through Humboldt's Parisian connections (chiefly François Arago, 1786–1853), Galle was awarded the Lalande Prize from the Paris Academy of Sciences (received on 2 November 1840 as a gold medal and a money prize of 635 Francs), and finally, through Schumacher, Galle received three Gold Comet Medals by the king of Denmark (23 September 1841). He also received a job offer to be an assistant to Johann Heinrich Mädler (1794–1874) at Dorpat (Tartu), which Galle, however, declined. (Wattenberg, 1963:43–44).

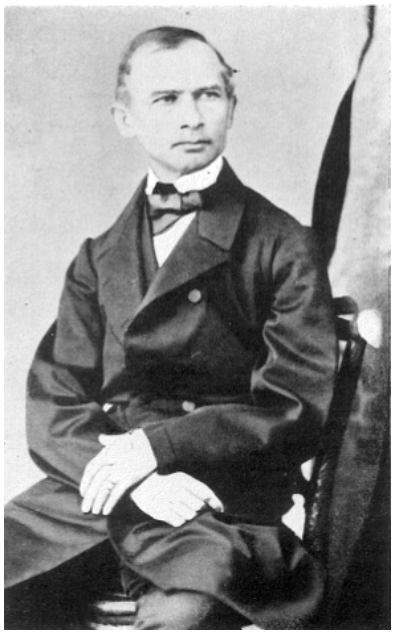

**Fig. 6.9** Johann Gottfried Galle (9 June 1811 – 10 July 1910). By the time of Neptune's discovery, Galle was internationally known as a discoverer of three comets in the same year. This photograph of Galle was taken when he was about 45 years old. (Credit: Wattenberg 1963:70)

Perhaps the most important benefit of the discovery of the three comets was that in June 1846 Galle was also awarded a stipend by the King of Prussia in the amount of 150 Thalers to cover the costs of the promotion to a Doctor of Philosophy (Wattenberg, 1963:44). Encke suggested the topic: a re-reduction of Ole Rømer's (1644–1710) observations called *Triduum observationum astronomicarum*. These observations were all that was left of Rømer's observation of more than 1,000 stars in 1706 after fire ravaged Copenhagen in 1728. They consisted of observations of 88 stars, the Moon, the Sun and all the known planets, as well as an observation of the lunar eclipse on 21 October 1706. Galle completed the work and submitted his thesis on 26 December 1844. On 2 January 1845 he paid the fee (100 Thaler and 9 Silbergroschen) and defended the thesis on 1 March 1845 (Wattenberg, 1963:45–47). The examination committee consisted of Wolfers, a co-worker at the Observatory, R. Jacobs (1807–1877), a professor at the Joachimsthalschen Gymnasium, and G. Michaels (1813–1895), an assistant at Friedrichs-Wederschen Gymansium.

The *Triduum* also contained three observations of Mercury in 1706. Galle was aware of Le Verrier's work on Mercury (Le Verrier 1843), in which he was collecting all observations of Mercury, current and old, in order to resolve its orbital anomalies (see Chapter 9.5). Galle, therefore, thought his "ancient" observations might be of help and sent Le Verrier a copy of his dissertation.

**6.6.2 Letters from Paris**

It is tempting to say that this Galle's action was a crucial moment for the discovery of Neptune in Berlin. The fact is that Le Verrier wrote a letter on 18 September 1846 to Galle (translation cited from Grosser, 1962:115–116, with the interpolation of orbital elements from p. 111):

> Paris, 18 September 1846
>
> Sir, – I have read with much interest and attention the reductions of Roemer's observations which you have been kind enough to send me. The prefect clarity of your explanations, the complete rigor of the results which you present are on a par with those which we should expect from a most able astronomer. At some future time, Sir, I will ask your permission to review several points which interested me, in particular the observations of Mercury included in your paper. Right now I would like to find a persistent observer, who would be willing to devote some time to an examination of a part of the sky in which there may be a planet to discover. I have been led to this conclusion by the theory of Uranus. An abstract of my research is going to appear in the *Astronomische Nachrichten*. I will then be able, Sir, to offer my excuses to you in writing, if I have not fulfilled my obligation to thank you for the interesting work which you sent me.
>
> You will see, Sir, that I demonstrate that it is impossible to satisfy the observations of Uranus without introducing the action of a new Planet, thus far unknown; and, remarkably, there is only one single position in the ecliptic where this perturbing Planet can be located. Here are the elements of the orbit which I assign to this body:
>
> | | |
> |---|---|
> | Semimajor axis | 36.154 |
> | Sideral period | 217.387 years |
> | Eccentricity | 0.10761 |
> | Longitude of perihelion | 284° 45' |
> | Mean longitude, January 1, 1847 | 318° 47' |
> | Mass | 1/9300 |
> | True heliocentric longitude, January 1, 1847 | 326° 32' |
> | Distance from the Sun | 33.06 a.u. |
>
> The actual position of this body shows that we are now, and will be for several months, in a favourable situation for the discovery.
>
> Furthermore, the mass of the planet allows us to conclude that its apparent diameter is more than 3" of arc. This disk is perfectly distinguishable, in a good glass, from the spurious star-diameters caused by various aberrations.
>
> I am, Sir, your faithful servant,
>
>                  U.J. Le Verrier
>
> Will you convey to Mr. Encke, although I have not had the honour of meeting him, my compliments and deep respect.

This is a remarkable letter, not only because it, in hindsight, initiated the epoch-making observations at Berlin. The letter starts as an acknowledgment, about year and a half late, full of praise and flattery. However, with the fifth sentence Le Verrier changes the topic completely, stating a desire that the "persistent observer" looks for his prediction on the sky. Why did Le Verrier remember Galle so suddenly and send him a letter? Why did Le Verrier expect that Galle would follow his request? The postscript is equally interesting: Le Verrier was not acquainted with Encke; was this the reason to choose Galle, who already wrote to him, instead of directly writing to Encke?

The Staatsbibliothek zu Berlin keeps a copy of a letter by Le Verrier to Schumacher dated on 1 October 1846, which started with (in French): "When I sent you my work, I wrote directly to Mr. Struve and Mr. Galle." (Le Verrier to Schumacher, 1 Oct 1846). Le Verrier was referring to his calculations that he presented at the Academy in Paris on 31 August, but he did not explain why he sent letters to Galle in Berlin and to Otto Wilhelm von Struve (1819–1905) at Pulkovo observatory. They were junior members of the respective observatories, although Otto Struve had already started taking over the duties of the observatory in the mid-1840s from his father Friedrich Georg Wilhelm von Struve. Wilhelm Struve took over the directorship of the Pulkovo Observatory close to St. Petersburg and moved there from Dorpat (Tartu) in 1839. Soon after he mostly devoted himself to science letting his son run the observatory. Following his father's death in 1864, Otto Struve became the official director of the Pulkovo Observatory.

Le Verrier's choice of astronomers asked to do the search can be traced to his correspondence with Schumacher. In particular, Le Verrier and Schumacher had already exchanged several letters regarding Le Verrier's theory. For example, in a letter dated 30 June 1846 Le Verrier (Le Verrier to Schumacher, 30 June 1846) informed Schumacher of his results on Uranus and prediction of the location of an external planet. In that letter, Le Verrier did not send any detailed calculations as he was working on improving them, but promised to send to Schumacher the final calculations as soon as they were done. Le Verrier's added: "… as the planet in question will arrive at its opposition, in the middle of August, I will ask for your benevolence for the immediate printing of my article."

Schumacher understood Le Verrier's reason for a speedy publication, and in the letter now surviving only partially printed in the *Centenaire de la naissance de U.J.J. Le Verrier* (and not dated; Centenaire, 1911:14) he gave advice as to who would be able to effectively search for the planet. It was clear to Schumacher that one needed a telescope of excellent quality and large aperture, and he suggested Le Verrier to contact the owners of the two biggest telescopes at the time: "M. Struve" and "Lord Rosse."

In 1846, the Pulkovo Observatory had the world's biggest refractor with a 38 cm aberration-free lens, built in Munich by Merz & Mahler, the successor of Fraunhofer's optical company (Mathematisch-Feinmechanische Institut), that also finished Fraunhofer's telescopes for Königsberg and Berlin. Schumacher was, however, not specific about whom to write, the father Wilhelm or the son Otto. The other suggestion referred to what was then the largest reflecting telescope in the world (until the Mount Wilson 100-inch saw first light in 1917). This was a telescope with a gigantic speculum-metal mirror of 1.8 m diameter, called the Leviathan of Parsonstown, that had been completed in 1845 by William Parsons (1800–1867) on his estate at Birr Castle in Ireland.

Le Verrier's letter from 30 June is remarkable for two other aspects. It was written two days after a letter he sent to Airy (Le Verrier to Airy, 28 June 1846) answering Airy's question regarding the radius vector of Uranus (see Chapter 4.10). Le Verrier considered the radius vector issue important enough to write about it to Schumacher as a proof that his calculation was indeed correct. He declared: "Thus, although I first occupied myself only with the longitudes deduced from the oppositions, the theory I arrived at did not fail to bring to the radius vector of Uranus the corrections required by the observations made in the quadrature."

This letter also entered the German literature about the discovery of Neptune in an indirect way. Schumacher sent it to Gauss as part of another letter (Schumacher to Gauss, 7 July 1846; Briefwechsel, 1863b:176) and asked Gauss's opinion on the matter. Gauss answered (Gauss to Schumacher, 11 July 1846; Briefwechsel, 1863b:179) by saying: "The weight of his [Le Verrier's] hypothesis cannot be judged until his calculations are submitted for substantive examination."

This statement was sometimes used to show how prominent astronomers of the time did not put too much trust in Le Verrier's prediction (e.g., Wattenberg 1946, Dick 1986), but it should be taken at a face value: Gauss simply had not seen Le Verrier's actual work.

Le Verrier did not follow the advice of Schumacher fully, for a good reason. For example, unlike the Pulkovo equatorial, Lord Rosse's reflector was a meridian instrument. Of course, there were other large telescopes, for example, in Cambridge. Airy did not answer Le Verrier's suggestions: *"If I could hope that you will have enough confidence in my work to seek this planet in the sky, I would hasten, Sir, to send the exact position to you, as soon as I have obtained it."* (Le Verrier to Airy, 28 June 1846).

Regardless, Le Verrier did directly ask O. Struve and Galle (there is no indication Le Verrier was aware that O. Struve was taking care of the Pulkovo Observatory's day-to-day work). If we are allowed to speculate, perhaps Le Verrier thought junior (and similar age – Le Verrier was born 1811, Galle in 1812 and O. Struve in 1819) astronomers might be more enthusiastic about the search, or he thought that a son (O. Struve) would be able to convince the director-father (W. Struve) to attempt the search. Le Verrier's choice in asking Galle was, however, very fortunate.

**6.6.3 Party crashers**

The letter from Le Verrier arrived in Berlin on 23 September 1846, the 55$^{th}$ birthday of Encke. As his plan was to skip the work that night and enjoy the evening party, Encke gave permission to Galle to conduct the search. Galle's own account (Galle 1877) was that Encke, who had expressed his "doubt and disproval" about a possible planet as the solution to Uranus problem, understood that Galle had "*eine gewisse moralische Verpflichtung zum Nachsehen an der betreffenden Stelle*" (a certain moral obligation to look up the position in question), i.e. to search for the planet now that he had received the letter from Le Verrier (Galle, 1877:350). Their conversation was overheard by a third person, Heinrich Louis d'Arrest, who was a student of Encke and a volunteer at the observatory, living in "a room under the roof" of a nearby building.

d'Arrest was native of Berlin, born in 1822, a descendant of a family of French Huguenot refugees (Dreyer 1876). He attended a local school and Collège Français in Berlin, and later enrolled at Berlin University. From the start he showed a special interest for astronomy and was called to assist at the Berlin Observatory. By September 1846, d'Arrest had distinguished himself by discovering a comet on 9 July 1844 (one already discovered two nights before in Paris), and then another comet on 28 December 1844, for which he was awarded the Gold Comet Medal from the King of Denmark. His work at the Berlin Observatory was, however, more theoretical as in 1846 he published his tour de force calculation of the orbit of the recently found asteroid Astrea in *Astronomische Nachrichten* (d'Arrest 1846).

As Galle reported in his first description of that famous observing night (Galle 1877), the sky was clear and conditions favourable. Galle's plan was to look for an object with a clear disk-like appearance as Le Verrier suggested it should be 3" in diameter, distinguishable by the Fraunhofer refractor, even in Berlin's unsteady atmosphere. d'Arrest was likely given the task of preparing the observations (Galle, 1882b:96), such as determining what position in the sky Le Verrier's prediction actually placed the planet, but at first neither Galle nor d'Arrest used more than the standard Harding's star chart that was in the dome (Fig. 6.2).

After about an hour of sweeping the estimated zone looking for a disk-like object, Galle was ready to abandon the project. The Fraunhofer refractor in Berlin was at that time one of the world's largest and best telescopes. Four decades later, Dreyer (1882) listed only five other observatories worldwide that at the time had a superior equatorial telescope: Pulkova, Harvard College, Munich, Cambridge, and Markree Observatory.[4]

---

[4] Dreyer's 1882 recollection was inaccurate concerning one telescope. Harvard's 15-inch (38-cm) Merz and Mahler refractor was not mounted until June 1847. In September 1846, the largest telescope in the United States was the 11-inch (28-cm) Merz and Mahler refractor at the Cincinnati Observatory (see Chapter 9).

## 6. "That star is not on the map": The German Side of the Discovery

In spite of the Berlin telescope's large aperture and superb optics, Galle and d'Arrest were not able to find a disk-like object. But they were fully aware that Harding's map was not complete. Galle (1877:351) reported that even though they were well aware of the existence of the Royal Academy of Berlin maps (in his second retelling of the events, Galle (1882a:219) also mentions that he had corrected some of them), and that they knew how the Hora IV map was used to find Astrea, Galle did not think there was another such chart that they could use. In that moment, however, d'Arrest remembered that there was a chart lying somewhere among Encke's papers that should cover the part of the sky they were examining.

Galle and d'Arrest went to Encke's hall (Vorzimmer), as Galle knew where the maps were stored, and indeed, they found Hora XXI by Carl Bremiker (Fig. 6.5). Back in the dome they divided the work: Galle was at the Fraunhofer looking for stars and describing their locations on the sky, while d'Arrest checked if the stars were drawn on the map at a nearby desk. At some moment after 22:00 local time[5] Galle reported a star that d'Arrest could not find on the map.

Did Galle say: "What?" Did d'Arrest say: "Can I have a look?" Did the two young men dance in the dome of the Berlin Observatory next to the sixth or seventh largest telescope in the world? Did they wonder how it was possible for Le Verrier to be so precise in his prediction, that the possible planet was scarcely more than a degree or so away from the predicted position? Maybe all this happened, or maybe nothing at all. What is certain is that Galle and d'Arrest turned from discoverers of a new planet to excited party crashers. The report (Galle, 1877:352) states that Encke, presumably still at his birthday party, was "given all details" and immediately left to join in with the observations. Expressing a bit more emotion Galle (1877:352) reported: "The evening of the 24. Sept had to be awaited with much excitement."

The reason was to determine the actual motion of the planet. Four hours of observations on the night of the 23rd were not able to determine with certainty that the object really moved against a backdrop of stars, although there was an indication. They also recognised the disk-like appearance of the planet, as predicted by Le Verrier, although the measured diameter was somewhat smaller than expected. Once the object was located, both Galle and Encke confirmed that the apparent size of the planet was about 2.7" or 2.9", compared to Le Verrier's estimate of a bit over 3". One can only imagine how long the day of the 24th was for the three observers, as they waited to confirm the existence of the planet.

The night of 24 September finally settled, but the skies were not as clear as on the previous night and the observations had to be soon called off due to clouds. Nevertheless, a motion of 4' in right ascension along the path of Le Verrier's prediction was sufficient to confirm that the object was a new planet. Next morning, and during following days, Galle and Encke were busy writing letters to announce their amazing discovery.

Galle sent the celebrated letter to Le Verrier dated 25 September, here presented at full length for the first time in a translation by James Lequeux (Galle to Le Verrier 25 September 1846; see also Chap. 5):

> Berlin, 25 September 1846
>
> Sir, - The planet whose position you had indicated <u>really exists</u> [Galle's underlining]. On the very day I received your letter I found an eighth magnitude star, which did not appear in the excellent chart *Hora XXI* (drawn up by Dr. Carl Bremiker) from the collection of celestial charts published by the Royal Academy of Berlin. The observation of the following day clinched the matter: here was indeed the planet we were looking for. We compared it, Mr. Encke and I, with the great refractor of Fraunhofer, with a ninth magnitude star (a) from Bessel Zone 119 21h 50m 51s.00 -13° 30' 7".9 and we have found:
>
> Mean Berlin time
> Sept 23. 12h 0m14s.6   Plan. = (a) + 21' 21.".5 in RA

---

[5] Encke's report of the discovery (Encke 1846), lists 22:52 for the first observation of Neptune with the precisely determined coordinates. There are three more observations at 23:47, 00:52 and 01:08, with Galle listed for the first three and Encke for the last one. It is however likely that the planet was discovered before these scientifically reported measurements.

```
                                        = (a) + 1  36.8   in Dec
          Sept 24.   8h 54m40s.9  Plan. = (a) + 20' 19".9 in RA
                                        = (a) + 1  15.9   in Dec
          Apparent place of the (a) star from Bessel
                    Sept 23   327° 58' 2".5  -19° 25' 49".0
                    Sept 24            2".4            49".0
```

We have compared twice this star with Piazzi XXI 944, whose position is also found in Taylor and hence in the British Association Catalogue. From these comparisons and the position in the Brit. Ass. Cat., one has for (a):

```
Sept 23   327 57 54.5  -19 25 45.0
  24            54.4         45.0
```

This determination will be preferable, Bessel being worried about Zone 119.
Thus we have for the planet:

```
Mean Berlin time         RA              Dec
Sept 23  12h  0m 14s.6   328° 19' 16".0  -13° 24'  8".2
Sept 24   8h 54m40s.9         18  14.3         24  29.7
```

Also, the diameter seemed near 3 seconds; however, one can only be sure in case of very favorable atmospheric circumstances, and it is mainly the chart that facilitated the search.

This planet might be worthy of being called Janus, one of the most ancient deities of the Romans, whose two-sided face would signify its position at the frontier of the Solar System.

J. G. Galle

A great irony of science is that while it took four days for a letter to travel from Paris to Berlin, it took 11 days to arrive at Pulkovo. Le Verrier's letter urging Otto Struve to look for the planet arrived on 29 September. By the time O. Struve was ready to observe two days later (due to bad weather), another letter arrived, this time from Encke in Berlin, announcing the discovery (Dick 1986). In his autobiography, O. Struve wrote that, as soon as he got the letter from Le Verrier and the observing conditions were good, he calculated the predicted position and started searching for the planet based on its disk diameter (O. Struve also pointed out that Pulkovo still didn't have the Bremiker map). In that moment his father W. Struve arrived with the letter from Encke and the coordinates of the discovered planet. O. Struve answered: "*Sag mir nichts nähers, ich will ihn auch unabhängig aufsuchen*" (Don't tell me anything, I want to find it independently). Within quarter of an hour he was able to call his father to the tower to show him the planet. (Dick, 1986:257).

Not everything was exactly as O. Struve wrote in his autobiography many years after the event. Wolfgang Dick (1986:257), based on Wilhelm Struve's correspondence, deduced that Encke's letter must have arrived by 1 October, while the first night suitable for observing in Pulkovo did not occur before 3 October. Nevertheless, Wilhelm Struve wrote that his son was able to recognise the planet by its disk, which is also evident from O. Struve's letter to Le Verrier published in *Centenaire de la naissance de U. J. J. Le Verrier* (Centenaire, 1911:23) where O. Struve states that the planet could be recognised even without a map by its clear disk, 2.78 seconds of arc in diameter, consistent with Berlin measurements.

**6.7 The Aftermath**

If Galle and d'Arrest were barely suppressing their excitement breaking up their director's birthday party, what was going through Le Verrier's thoughts when he received the letter from Galle four or five days after it was sent from Berlin on 25 September? Did his hand tremble while he was cutting the envelope? He read the opening sentence of Galle's letter "*Le planète, dont vous avez signalè la position, rèellement existe,*" which changed his life.

Le Verrier answered Galle as soon as he received the announcement (Le Verrier to Galle, 1 October 1846):

> Paris, 1st October 1848
>
> Sir, - I thank you heartily for your alacrity to tell me about your observation of 23 and 24 September. Thanks to you, we are definitively in possession of that new world. The pleasure I have felt when seeing that you have encountered it at less than a degree of the position I had given, is a bit darkened by the idea that if I had written you four months earlier we would then have obtained the result we have just reached now.
>
> I will communicate your letter next Monday to the Academy of Sciences.
>
> Let me hope that we will frequently go on with a correspondence that is starting under such happy auspices.
>
> I am with my high consideration, Sir, your devoted servant.
>
>       U.J. Le Verrier
>
> Please be kind to write me rue Saint-Thomas-d'Enfer, n° 5. I do not belong to the Observatory, where you have sent your letter.
>
> The Board of Longitudes here has proposed the name of Neptune, with a trident as symbol. The name of Janus would mean that this planet is the last of the Solar System, something that there is no reason to believe.

This letter, preserved among Galle's papers and kept in the Staatsbibliothek in Berlin, shows Le Verrier's happiness on the discovery, the evidence of an anxiety (Galle's announcement was sent to a wrong address) as well as the foretelling of the trouble that will engulf him, as Le Verrier was starting his battle to name the predicted planet.

The lives of Neptune's telescopic discoverers changed as well, albeit in different ways. The attention for the discovery of Neptune was focused on the predictors, Le Verrier and Adams. In spite of this, both Galle and d'Arrest became internationally recognised astronomers with respected careers, as directors of observatories and professors at universities.

It is, however, surprising that d'Arrest's role in the discovery was not widely known until after his death in 1875. The first two encompassing records of Neptune's telescopic discovery present it in mutually consistent ways, which are, however, notably different from that written by Galle in 1877 and accepted today. John Pringle Nichol (1804–1859) in his book published in 1849 *The Planet Neptune: an Exposition and History* wrote:

> Now, it so happened that the map of that precise region where the new planet was expected, had been completed by Dr. Bremiker; and it was printing, or just printed at Berlin—I believe that the Observatory of Berlin had obtained the proof-sheet. The Astronomers of this Institution were thus in a position of power regarding such inquiries, enjoyed by no other Observatory in existence: they had simply to notice Bremiker's Map and then the Sky— observing if there was a discrepancy between the two pictures, that could be accounted for by the planetary motion of some one star: so that—with their renowned sagacity, and the excellence of their Instruments—an inspection of the Heavens on one clear night might accomplish the resolution of this great problem. And thus it even was; the Planet was discovered actually by M. GALLE, on the very evening of the day on which he received the letter of LEVERRIER indicating its place. (Nichol, 1849:88–89)

A year later, Bernhard von Lindenau (1780–1854) devoted only a single sentence to the discovery in his 30-page essay printed in the *Astronomische Nachrichten*:

> As Le Verrier's communication of his newest location of the presumed planet with the desire of its search arrived on 23 Sept 1846 to Dr. Galle at the Berlin Observatory, its discovery succeeded in that same night, as near the calculated place a star of 8th magnitude was perceived, which was missing on the Dr. Bremiker's academic star map and the next evening it was recognised as the sought for planet, by the motion in the sense of Le Verrier elements. (von Lindenau, 1848:12)

Both accounts mention only the astronomer Galle and highlight the availability of Bremiker's map. The search is presented as a matter of pointing the telescope, and d'Arrest is not mentioned at all. The events in Berlin did not spark a scandal, but rumours were going around that not all was as it seemed.

Understanding the roles of Galle, d'Arrest, and Encke, as well as the pivotal contribution of Bremiker's star map, is a necessary ingredient of any complete account of the discovery of Neptune.

### 6.7.1 Johann Gottfried Galle

Among two discoverers and three observers in the dome of Berlin Observatory in the night of 23 September, Galle received the most attention. Already on 4 October 1846, Galle was decorated with "Roten Alderorden IV. Klasse" by King Friedrich Wilhelm IV of Prussia. This was soon followed by the news that the French king Louise-Philippe had named him the Knight of the Legion of Honour. The Danish king, who had already awarded three gold medals in 1840 to Galle for discovering comets, sent him another gold medal. Galle's award season finished on 31 July 1847 when he was decorated with the "Roten Alderorden III. Klass," but more importantly, on Humboldt's initiative Galle's salary increased by 200 Thalers per year in October 1846 (Wattenberg, 1963:53–54).[6]

Collecting the figurative laurel wreaths, Galle withdrew from discussions about Neptune. In the announcement letter to Le Verrier he suggested the name of the planet (Janus), but as Le Verrier refused it as a new name had already been chosen by the Bureau des Longitudes (Neptune), Galle decided he should recognise Le Verrier's "right of the discoverer"' to name the planet and focused on his usual work at the observatory. At that time, he was also collaborating with Humboldt, providing information and calculations (Wattenberg, 1963:65–69) for Humboldt's epoch-making multi-volume *Kosmos*. In 1848 came a possibility for leaving Berlin. Galle was a candidate for the directorship of the Königsberg Observatory that had been vacant since Bessel's death in early 1846. The Neptune fame, however, did not help him at that time, and, on the insistence of Carl Gustav Jacob Jacobi (1804–1851), the post was given to Bessel's long-standing assistant, August Ludwig Busch (1804 –1855; Wattenberg 1963:58). Jacobi's point was that Galle might be known for discovery of a planet and four comets, but Hencke discovered two planets (Astrea in 1845 and Hebe in 1847) and "*Comet discovery is the job of every dilettante who knows the starry sky and has therefore become the business of several ladies*" (Knobloch 2013:59).

Galle travelled to Frombork (Frauenburg) in 1851, the town where Copernicus had lived and worked, to observe the total solar eclipse of 28 July, and, then moved to Breslau (Wrocław), finally getting his tenure as the director of the observatory there and a professor at the university. Galle married and had a scientifically fruitful 44 years as the director. He investigated the climate and the Earth's magnetic field and proposed a novel technique to estimate the solar parallax using minor bodies of the Solar System (Galle, 1873). He thereafter coordinated a network of observatories to observe the solar parallax by his technique, and in 1875 published their results (Galle, 1875). Galle retired aged 83, on the 50th anniversary of his PhD thesis. In 1835, just after moving in the Berlin Observatory at the start of his astronomical career, Galle observed Halley's comet. He had the good luck to do it again in 1910, just a few months before his death. (Wattenberg, 1963:106).

Neptune featured little in Galle's life, except at two periods of the post-discovery career. The pleasant events happened towards the end of his life, when Galle was celebrated as the discoverer. He received accolades in 1896, for the 50th anniversary of the discovery. (Wattenberg 1963:109–110). This was followed with the award of "Stern zum Kronen-Orden 2. Klasse" for his 90th. (Wattenberg 1963:114). Somewhat less pleasant events happened some 30 years after the discovery, when Galle felt the need to come back to the discovery of Neptune with no fewer than three published statements.

---

[6] Another person that got awarded for the discovery of Neptune was Carl Bremiker. On personal request by Encke, Bremiker was awarded with the Gold, Silver and Bronze Leibniz Medals of the Royal Academy of Berlin: (Bruhns, 1869:122–123).

They form the important evidence for the actual role the Bremiker map played in the discovery, but, crucially, two of the letters also provide authoritative evidence for d'Arrest's role in the discovery.

**6.7.2 The Star Chart**

The significance of the Hora XXI map in finding the new planet, and its curious, but, as we have seen earlier, understandable location at the Berlin Observatory, induced early historians of the event to conclude that its existence was the reason why Le Verrier wrote to Galle. That supposition, however, can be invalidated easily. Before we do that, it is instructive to see why the star chart was considered so important.

On several occasions immediately after the discovery, Encke emphasised the crucial role of the map: in his first report of the discovery to *Astronomische Nachrichten* (Encke 1846), the congratulatory letter to Le Verrier (Centenaire, 1911:20–22), as well as in the address to the Royal Academy of Berlin, also printed in *Astronomische Nachrichten* (Encke 1847). In the letter to Le Verrier, Encke was very specific (translation from Lequeux, 2013:37): "Without the fortuitous circumstance of having a chart containing all the fixed stars down to the tenth magnitude [of this particular area of the sky], I do not believe the planet would have been found."

Encke, in the first official scientific report, implicitly suggested that Le Verrier's idea of recognizing the planet by its disk didn't work, but having a map allowed for a quick discovery:

> By a letter that arrived here on 23$^{rd}$ Sept. Mr. Le Verrier asked Dr. Galle especially to look after it [the planet], probably led by an assumption, pronounced in his treatise, that the planet will be identified by a disc. The same evening Mr. Galle compared the excellent map, drawn by Mr. Dr. Bremiker (Hora XXI of the academic star charts), with the sky and noted immediately that very close to the position determined by Mr. Le Verrier a star of the 8$^{th}$ magnitude was missing in the map. (Encke, 1846:49-50)

Encke was similarly convinced when speaking to the academicians in Berlin that the discovery was due to the existence of the map (Encke 1847:191): "The new planet whose orbit and location were announced by Mr. Le Verrier in Paris in advance, was almost immediately found by Mr. Dr. Galle on the 23rd of September at the local observatory, where the Hora XXI of the academic star charts [...] was compared with the sky…"

Here are those two remarkable aspects found in Lindenau's, Nichol's, and other accounts: Encke mentioned only Galle as the discoverer and that the planet was found "almost immediately" by comparison with the star chart.

Such impressions certainly fixed the importance of Bremiker's map, together with comments, chiefly among British astronomers, that its absence in Cambridge hindered Challis's search (e.g. Nichol, 1849:131). It is, therefore, not surprising that this notion started to spread throughout the literature.

The only authentic retelling of the events leading to the discovery was not available until 1877 (Galle, 1877). By then, Galle was the only person closely related to the discovery of Neptune in Berlin still alive. Encke died in 1865 in Berlin, d'Arrest in 1875 in Copenhagen and, on the day of the 31st anniversary of the discovery, Le Verrier died in Paris. In writing his own account of the events, Galle had three goals: to describe the discovery as he saw it happened, particularly now that he was the only surviving member of the observing team, place the due importance to Bremiker's map, and highlight the contribution of d'Arrest (see Section 6.7.2). He was prompted by the publication of the translation of Hugo Gyldén's (1841 – 1896) book, written originally in Swedish, *Die Grundlehren der Astronomie*:

> When the news of Paris arrived at the Berlin Observatory, the then assistant to the observatory, Mr. d'Arrest, hurried to design a small map of the designated area of the sky to facilitate the search for the planet. As soon as he had pointed the telescope to the relevant celestial region, the observer Dr. Galle arrived, looked into the telescope and - saw the planet. (Gyldén, 1877:248)

Gyldén's account of the discovery is obviously a somewhat corrupted and much shortened retelling of the story that was by then circulated. Its origin, as it will become evident later, might have been a rumour. It is remarkable in two aspects: it mentions d'Arrest prominently, and it actually does not mention which map d'Arrest used.

Galle's 1877 retelling forms the current framework of the story (Wattenberg 1946; Grosser 1962; Standage 2000) as described in Section 6.6.3. The English language literature was, however, somewhat slower in accepting Galle's version. Smart (1946) wrote: "A young student-observer, d'Arrest, suggested that the first thing to do would be to find out if Bremiker's star-chart (Hora XXI) had been finished." He rightly put d'Arrest in the center of action, but simplified the events in a similar manner as Gyldén's book. Other additions are common: for example, Grosser (1962:118) gave a large role to d'Arrest, suggesting even that he was the one who went to inform Encke, although there is no evidence for this. Standage (2000:119) incorrectly considered the Berlin sky charts as volume in an atlas that "had recently become available, … and had yet to be widely distributed." On the other hand, the French summary of the events (Danjon, 1946) was shorter, stating that following the suggestion of a young assistant d'Arrest, Galle used the Berlin Academy map. This account is essentially correct, but it leaves out the subsequent role of d'Arrest.

Galle's statement in the *Astronomische Nachrichten* might have rehabilitated d'Arrest's role, but it did not stop the speculation about the importance of the map. According to the German translation *Populäre Astronomie* of the book *Popular Astronomy* by the American astronomer Simon Newcomb (1835–1909) (Newcomb 1881:392): "Le Verrier wrote to Galle in Berlin and asked him to look for the planet based on the just finished page of the 21h in the academic star chart."

This statement actually was an addition to the text by the editor (Rudolf Engelmann, 1841–1888) of the German edition, and it does not appear in the English original. The British edition from 1878 states only:

> Leverrier wrote to Dr. Galle, at Berlin, suggesting that he should try to find the planet. It happened that a map of the stars in the region occupied by the planet was just completed, and on pointing the telescope of the Berlin Observatory, Galle soon found an object which had a planetary disk, and was not on the star map. (Newcomb, 1878:361)

Here again, d'Arrest is not mentioned, while it is not clear whether the planetary disk had been helpful in the search (contrary to Encke's reports).

The statement in *Populäre Astronomie* prompted Galle to issue a rebuttal in *Astronomische Nachrichten* (Galle, 1882a:221) stressing: "*Von dieser Karte hatte jedoch Le Verrier […] keinerlei Kenntniss*" (Of that map Le Verrier […] had no knowledge). Crucially, Galle and d'Arrest did not prepare for the search with Berlin charts in mind. That was already explicitly stated in Galle's (1877) account of the observations, after it became clear that searching for the disk alone did not produce results (Galle, 1877:351): "…one had to think about the procurement of a star map, for which there were, next to the Harding's Atlas, only the Berlin Academy star charts, still very full of gaps and long awaiting completion."

Galle's (1882a) letter to *Astronomische Nachrichten* did not seem to have a big influence on the German publisher Wilhelm Engelmann; the second edition (1892) of the *Populäre Astronomie*, even though the editor has changed, kept the same sentence (Newcomb 1892:403). Only in the third edition in 1905, the editor Hermann Carl Vogel (1841–1907), removed the explicit statement that Le Verrier asked Galle to use the map (Newcomb, 1905:413). It is important to point out that the same publisher, Engelmann, published both Gyldén's and Newcomb's books, which might explain the propagation of the error. A similar coincidence might be that Vogel was the director of the Astrophysikalischen Observatorium Potsdam between 1882–1907, coinciding with Galle's move to Potsdam after retirement.

Following Galle's statements, it can be assumed that Hora XXI, while ready and waiting to be distributed, was not included among the standard material for assisting observations at the Berlin Observatory. That must be the reason why Galle and d'Arrest did not even consider starting the search using Hora XXI.

Nevertheless, even in Galle's German obituary, Wilhem Foerster (1832–1921) suggested that Arago, always in a close contact with Alexander von Humboldt, was undoubtedly aware of the availability of that particular map, and suggested that Le Verrier contacts Berlin: "… [Arago] undoubtedly had knowledge of the existence of such maps in Berlin and was thus able to take very good reason to point out the participation of Berlin to Le Verrier." (Foerster, 1911:20).

That is not exactly contrary to what Galle was trying to refute in Rudolf Engelmann's translation of Newcomb's book. Le Verrier's letter to Galle did not mention the maps in any way, but that does not mean that Arago or Le Verrier were not aware of the whole map project. The discovery of Astrea in 1845, using Hora IV map, happened sufficiently early for the general usefulness of the maps for discovery of new sky objects to be widely known. The insinuation that the existence of the map might have been related to the reason Le Verrier wrote to Galle can, however, be contradicted with the following three facts.

1) Le Verrier also wrote to O. Struve, who was not involved in the making of maps in any way. He might have acted on a suggestion by Schumacher to select people with best telescopes.
2) In July and August, when the search for Neptune started in Cambridge, there was no effort to use the Argelander's Hora XXII map (Fig. 6.10), completed in 1832 and present in Cambridge, which covered the region of Neptune's orbit for those months. There were several good reasons why Challis did not attempt a search with a map (see Chap. 4), but there are no reasons that Parisian astronomers were more confident in (or thinking of) using Berlin maps than their peers in Cambridge.
3) Le Verrier's letter, both to Galle and O. Struve, focuses on the fact that the planet should be discerned as a large (more than 3" in diameter) disk. A good telescope, and both Pulkovo and Berlin had first class telescopes (which was not the case for Paris), in Le Verrier's estimate should have been able to resolve it.

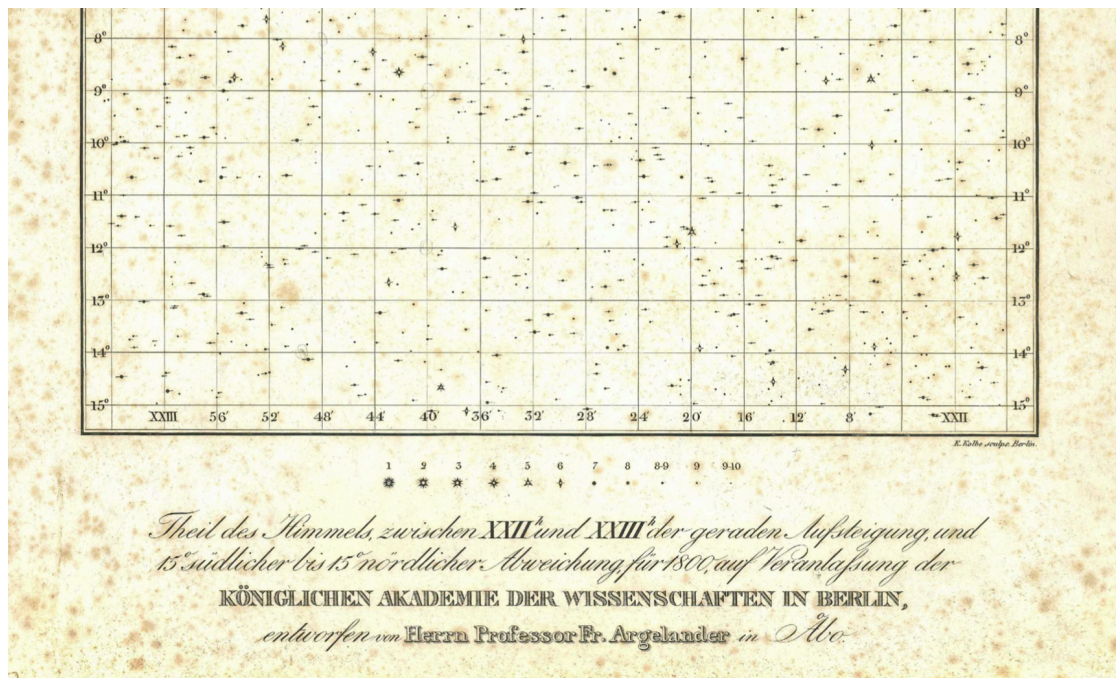

**Fig. 6.10** Hora XXII of the Royal Academy of Sciences in Berlin completed by Friedrich Wilhelm August Argelander in 1832, kept in the Library of Leibniz-Institut für Astrophysik Potsdam. In August 1846 Neptune could be located by using this map as it was passing through the rightmost quadrants (between 21h 56m and 22h) at the declination of approximately -13º (See Fig. 6.5 for comparison). A copy of this map existed in both Paris and Cambridge in 1846 (Credit: Library of Leibniz-Institut für Astrophysik Potsdam).

Le Verrier's letter is that of a convinced theoretician: it suggests that an observer equipped with a good telescope only needs to look in the direction indicated by the orbital elements, and the planet will be recognised by its large disk. In reality, such observations are typically more difficult than that, as Galle's and d'Arrest's search testified. (W. Struve, however, claimed his son recognised the planet by its disk-like appearance alone, Dick, 1986:257). Le Verrier, not trained as an observer, might not have fully grasped what it takes to find an unknown object in the sky. As a theoretician he understood the value of a good, aberration-free telescope, and that is most likely the reason he wrote to Berlin and Pulkovo, known for good telescopes. We might not know Le Verrier's reasons, but we can be certain that although Bremiker's map Hora XXI played a crucial role for the discovery, the search was not prompted by existence of the map.

**6.7.2 Heinrich Louis d'Arrest**

The role of d'Arrest in the discovery of Neptune was known only by a few until John Louis Emil Dreyer (1852 – 1926) wrote d'Arrest's obituary in German language. (Dreyer, 1876). That omission, however, did not hinder d'Arrest's career. Two years after the discovery, d'Arrest moved to Leipzig to a newly founded observatory under August Ferndinand Möbius (1790–1868). There he got a doctorate in 1850, published a treatise on the known asteroids in 1851, and married a daughter of his director in 1857.

In 1851, d'Arrest was invited to move to Washington D.C as an extraordinary professor (Dreyer, 1876, page 5), but the Leipzig University retained him by opening a professorship post for him. In the same year, he finished the Hora III of the Berlin star maps that Encke had entrusted to him some time earlier. In 1856 he was invited by the Imperial Russian government on O. Struve's suggestion to take over the directorship of the Kiev Observatory (Dick, 1986:251). After much deliberation, d'Arrest declined, but by 1857 he decided to move and relocate his family to not-so-distant Copenhagen, where he became the director of the observatory and a professor at the university. d'Arrest discovered a periodic comet 6P/d'Arrest (1851) and the asteroid 69 Freia (1862). His later work focused on nebulae and galaxies (of which he published a catalogue and for which he was awarded a Gold Medal of the Royal Astronomical Society) including pioneering work in spectroscopy of nebulae.

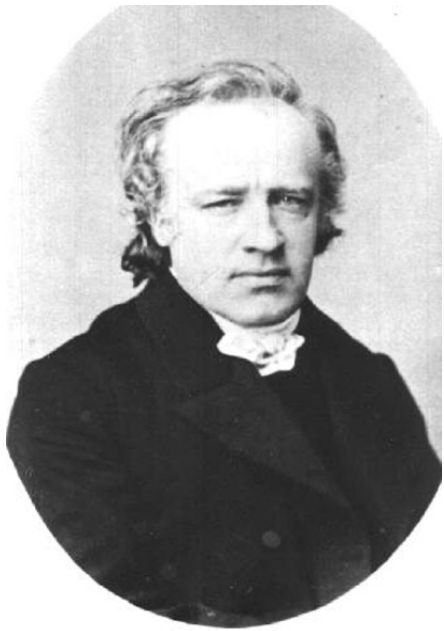

**Fig. 6.11** Heinrich Louis d'Arrest (13 August 1822 – 14 June 1875). His role in the discovery of Neptune became generally known only after his death, through popularisation by J. L. E. Dreyer and Galle's accounts. The photograph is by Bertel Christian Budtz Müller and is kept in Royal Library, Copenhagen (Wikipedia, Public Domain).

Dreyer, a native of Copenhagen, started frequenting the observatory even before he enrolled at the university in 1869 (Obituary Notices 1927). In 1870 he got a key to the observatory and started using it regularly, sometimes observing with d'Arrest, his professor of astronomy. d'Arrest seemed to enjoy talking about the discovery night to his students. Dreyer noted that d'Arrest said in one of his lectures "that he had been present at the finding of Neptune and that 'he might say it would not have been found without him.'" (Dreyer, 1882:64).

The first public announcement of d'Arrest's contribution is a short mention in Dreyer's obituary of d'Arrest (Dreyer 1876:3), where he wrote: "From 1846 he [d'Arrest] took part in the ongoing work of the Berlin Observatory, and acted as Galle's assistant when he, on September 23, 1846, found Neptune."

Nothing else was said in the obituary about Neptune's discovery. Nevertheless, Dreyer's statement and the description of the discovery night in the German edition of Gyldén's book, which mentioned the map and accentuated the role of d'Arrest, must have been sufficient to prompt Galle to write his version of events on the discovery night (Galle, 1877). Galle gave a relatively prominent role to d'Arrest, describing him as a student and an assistant ("Gehüfe") who worked at the Observatory (Galle, 1877:350). To make his work more practical and allow him to take part in the observations, he lived in an attic room of an adjoining building on the premises of the observatory. Galle continues that d'Arrest overheard the conversation between Galle and Encke (Galle, 1877:350): "When I communicated the arrival of Le Verrier's letter, the same [d'Arrest] expressed the wish to be present at the evening of the search for the planet, which, as was understood by itself, was gladly fulfilled."

In Galle's retelling, d'Arrest was mentioned once more, a few sentences later (Galle, 1877:351): "… it was first d'Arrest who raised the question of whether there should be taken a look at the academic star charts, if perhaps the position in question was already included among them."

This sentence is remarkable as it clearly highlights d'Arrest as the person who got the idea to look for Bremiker's map, but the phrasing is such that there is no doubt who was in charge in the dome of Berlin observatory on that night.

Galle's version of the events did not fully satisfy Dreyer, but nothing happened for five years. In the meantime, Dreyer started an international journal of astronomy, called *Copernicus*, and Galle again felt he needed to issue a statement regarding the latest incorrect description of the events, this time in *Populäre Astronomie*, the German version of Newcomb's *Popular Astronomy* (Galle, 1882a), as mentioned earlier. Dreyer did not miss the opportunity to write a short "Historical note concerning the discovery of Neptune" in his journal *Copernicus*. His motivation is probably best seen in this sentence (Dreyer, 1882:64): "As various conflicting accounts seem to have been circulated from time to time (without any having been printed except the one in Prof. Gyldén's book) this does not appear to me altogether useless."

Dreyer was not directly saying if the account in the German edition of Gyldén's book was correct, but here we have for the first time a clear indication that rumours were circling and that the editors of Gyldén's book knew them. Dreyer focused on what he heard directly from d'Arrest when they met and observed together:

> ... and [d'Arrest] continued: "We then went back to the dome, where there was a kind of desk, at which I placed myself with the map, while Galle, looking through the refractor, described the configurations of the stars he saw. I followed them on the map one by one, until he said: and then there is a star of the 8th magnitude in such and such a position, whereupon I immediately exclaimed: that star is not on the map!" (Dreyer, 1882:64)

Dreyer's mission was to establish:

> That d'Arrest thus not only first thought of looking for a map (without which the search might have proceeded as slowly as the operations at the Cambridge Observatory did), but actually *took part in the observation*, does not appear to be without historical interest, and it seems only just to that afterwards distinguished astronomer to say that Neptune was found by Galle and him, observing together. (Dreyer, 1882:64)

Dreyer's "Historical Note" was dated February 1882, and Galle saw it as part of the March issue of *Copernicus* (Galle, 1882b:96). He immediately sent a note "*Ueber die erste auffindung des planeten*

*Neptun*", dated 14 April 1882 to the same journal, which was published as part of the same volume as Dreyer's note (Galle, 1882b). Galle started with an explanation that he was writing to correct previous omissions of both Encke and himself. The first point Galle addressed was that Le Verrier wrote to him based on the "*frühere Correspondenz*" (previous correspondence).

It is important to stress that in 1882, Le Verrier's letter to Galle on 18 September 1846 was yet to be published. Neither was his connection to Le Verrier prior to the discovery generally known. Le Verrier's letter was first published, in German translation, in Julius Scheiner's (1858–1913) article, dealing with a wider issue of the "astronomy of the invisible" and including the discovery of Neptune as an example, in a general audience magazine *Himmel und Erde* (Scheiner 1893). Notably, Scheiner wrote that:

> …as with most important discoveries, and this one included, a number of legends and distortions got intertwined with them, so that in almost no popular work there is a correct representation of them; we therefore considered it interesting to give a strictly correct account of the situation here.

The transcript of the French original letter was presented in W.W. Payne's journal *Popular Astronomy* (See 1910: 475), then in Galle's obituary in the *Monthly Notices of the Royal Astronomical Society* (Turner, 1911: 275), and, with a number of other relevant letters, in the *Centenaire de la naissance de U. J. J. Le Verrier (Centenaire, 1911:14)*, which had a wide impact.

Returning to Galle: Galle's next point in his 1882 letter to *Copernicus* was to repeat the part of the 1877 retelling that Encke allowed him to attempt the search on his own. The crucial part of the text (translated from the original German) then followed:

> In the meantime, d'Arrest, hearing this, expressed the implicit wish to be allowed to participate in the observation. Although it was not my intention to relinquish this search and the possibility of an eventual observational success to another observer, such thoughts were remote and it would have seemed unkind to me, to in some way reject the wish of this young zealous astronomer, so I gladly gave my consent to the attendance. The same had, therefore, helped with writing down, looking up at the map, and perhaps a few other tasks, which I have no particular recollection of, which, however, as does not need to be discussed further, could have been carried out by me without a considerable loss of time, and while being personally appreciated, they were objectively irrelevant. On the other hand, I have always considered, as a significant contribution to a faster exploration of the planet, d'Arrest's quick memory of the academic maps, although here, too, opinions may and may have differed, to what extent it was inevitable and necessary for me to remember the academic maps.

Galle finished his letter repeating his regrets that he didn't highlight d'Arrest's contributions earlier.

This 1882 note was the last time Galle wrote of the discovery night, and it is clear that he was aware of the rumours. Perhaps to address those, he explicitly stated that d'Arrest did at least take part in preparations for the observations. It is easy to imagine that Galle, once he accepted d'Arrest as his co-observer, also told him to prepare what needs to be prepared for the night: e.g., convert Le Verrier's orbital elements to a location on the sky where they had to point the telescope. There is no doubt that Galle did not need d'Arrest to do that, but it is not surprising that a junior member of the team had done the (basic) groundwork. Grosser's (1962:117) statement that Galle "had computed the geocentric coordinates of the planet from Le Verrier's elements" is therefore likely not true. It was, most likely, done by d'Arrest.

Luck was also involved in the role d'Arrest played in the discovery. It might have happened that both Galle and d'Arrest were looking through the refractor searching for Le Verrier's disk, and that at some point even d'Arrest had a possibility to see the planet first. Perhaps, as Dick speculated, d'Arrest saved the day by remembering the map, when Galle was about to call off the search for Le Verrier's disk. Regardless of details and conjectures, d'Arrest's role is clear: he *was* part of the team of two, and thus there is no reason not to call him a *co-discoverer*.

Modern retellings of the Neptune discovery (e.g., Grosser, 1962:115–118; Standage, 2000:118–120) typically mention d'Arrest's role, although his name seems to be the first to be removed when a shorter version of the discovery is needed. (Dunkely, 2018:50). Furthermore, d'Arrest doesn't really feature in a prominent role even of longer accounts. This habit of mentioning d'Arrest in passing, while giving more space to Galle (or Le Verrier and Adams, the true stars of the event), is most plainly noticeable in how the scientific community honours the principal actors of the Neptune story. Adams, Le Verrier and Galle are names of craters on the Moon and Mars. The name d'Arrest was given to a crater on the Moon, and to a crater on Phobos, the satellite of Mars. More to the point, the ring system of Neptune, established by Voyager, is differentiated between rings named after (in order of distance from the planet): Galle, Le Verrier, Lassell, Arago and Adams (de Pater et al., 2018)[7]. We are left to hope that Neptune ring system is richer by at least one ring, and d'Arrest joins the illustrious company.

To summarise, Galle (1877), Dreyer (1882), and Galle (1882a,b) provide the description of the discovery night. They corrected common misconceptions from that time that are sometimes present in later recounts. They provide evidence that the Bremiker map was not the reason Le Verrier contacted Galle, that there were two observers and that Galle was given the responsibly for the search and leading the observations. The second member of the team, d'Arrest, had assisted in preparing the observations and provided crucial input by timely remembering the existence of the Hora XXI star chart. The planet was discovered through teamwork: Galle looking at the sky, d'Arrest looking at the map.

**6.7.3 Not everything is rosy in Berlin**

Dick was the first in modern literature to point out the puzzling lack of mention of d'Arrest in immediate reports of the new planet (Dick, 1986). His observation was prompted by the discovery of an autobiographical manuscript of Otto Struve at the Kharkiv Observatory in Ukraine (Dick 1985). At one point in it, O. Struve reminisced about the time when he was trying to bring in d'Arrest as the director of the Kiev Observatory. Around the same time (1856), O. Struve travelled to Berlin and met with Encke and wrote:

> d'Arrest gladly agreed [to come to Kiev], and when I came to Berlin with this news, I was surprised by a visit from Encke, who embraced me by saying casually: "You took a heavy weight from my heart by placing d'Arrest in such a favourable position; I have constantly been remorseful that I attributed the actual discovery of Neptune solely to Galle and, if possible, in his favour, while the principal merit of the discovery is due to d'Arrest" (translated from Dick 1986:251).

Struve's account also confirmed a rumour that was circulating. He wrote in his autobiography:

> In these words, I can only recognise a confirmation of the general rumour of the time: Galle had no confidence in Le Verrier's calculations and had made no preparations for finding the planet in the sky, but d'Arrest had done so, calculated the position in advance and prepared everything for the first favourable evening, to make a survey of the area of the sky in question, by comparing it with the first draft of Bremiker's map. Galle had approached [the telescope], and, because d'Arrest was already a little weary, proposed to temporarily replace him at the refractor. It coincidentally happened that in one of the first objects that Galle had registered in the sky, d'Arrest, who had compared Galle's information with that map, had to declare that this object was not on the map. With that, the planet was discovered and in that it must be admitted that Galle was the first to have seen the planet as such in the sky. (Dick, 1986:253)

This is a fascinating discovery by Wolfgang Dick, as it bears a close resemblance to what the German editor of Gyldén's book added about the discovery of Neptune. It contains two aspects that Galle tried to refute in his published versions of the event: that d'Arrest was the main observer and that the map was at hand.

---

[7] For the sake of completeness, the four illustrious names are also given to main belt asteroids, 1996 Adams (1961), 1997 Le Verrier (1963), 2097 Galle (1953) and 9133 d'Arrest (1960), where the number in parentheses is the year of discovery.

Struve's autobiography, however, did not stop here, as it mentioned two other interesting points. It first pointed out that Dreyer's version of the events from his 1882 "Historical Note" was met with some disapproval. Struve wrote: "Later objections to this conception were made, and [Anders] Donner [1854–1938] protested by claiming the merit of the discovery, based on d'Arrest's own statements to him, exclusively to Mr. Galle." (Dick, 1986:255). Nevertheless, O. Struve had a chance to ask d'Arrest personally how much of the rumour was true when they met in 1869. D'Arrest answered enigmatically that (Dick, 1986:255): "…there was some truth to the rumour, but he wished to admit that Galle deserved the main merit of actual discovery."

Encke's omission to mention d'Arrest created the possibility for all sorts of rumours and gossip. We might speculate if it had even contributed to the departure of d'Arrest from Berlin by 1848, even though Encke and d'Arrest's relationship did not seem to suffer (e.g., d'Arrest took over the work on one of the star charts). Our chief evidence about the rumour comes from an autobiography written some 40 years after the discovery. One should also consider that O. Struve was "robbed" of the possibility to search for the planet, as Le Verrier's letter arrived at Pulkovo on 29 September, while the announcement of the discovery from Encke arrived on 1 October, before Struve's observations could even start due to the weather.

Nevertheless, the source of the rumours was likely to have been, at least in their initial form, d'Arrest himself. After all, this is how Dreyer found out d'Arrest's role in the discovery. Worthy of a remark is the fact that d'Arrest lived for 10 years in Leipzig (where he was one of the two main astronomers). The publisher of Gyldén's and Newcomb's books, Wilhelm Engelmann (1808 – 1878), and his eldest son and the editor of the book Rudolf Engelmann, that first publicised d'Arrest's role, had their business also in Leipzig.

Encke's omission might have been accidental, in the spirit of time, where directors would report discoveries of their assistants, not necessarily mentioning their names. Encke did the same thing with Galle (Encke, 1837) when he reported in passing Galle's discovery of Saturn's C ring. This omission led to Galle's publications in 1851 and 1882 (the latter is the same paper where he points the mistake regarding the discovery of Neptune in the German edition of Newcomb's *Popular Astronomy*) about his role in the discovery of the ring before George P. Bond (1825–1865) at the Harvard College Observatory (Galle, 1851, 1882a). To this extent one should add that Encke, while a strict director who expected prompt execution of tasks, also allowed to his assistants "a free scope in the choice of theoretical studies, accounting tasks, observations and other practical examinations" (Bruhns 1869:191).

Everything was not rosy in Berlin immediately after the discovery, as is evident in an exchange of letters between Schumacher and Le Verrier. The Saatsbibliothek zu Berlin keeps Le Verrier's letter dated 7 January 1847, in which Le Verrier wrote:

> I am distressed to think that one might not be more reasonable in Berlin than in Paris, and that Mr. Galle might well have his part of inconvenience, as you were kind enough to write to me in confidence. It seems to me at present that we can no longer hope to receive an answer from Berlin. I suppose that Mr. Encke may have thought that all the merit was in having the Map built and that it was to him that the honour had to be returned to a large extent. There can be some truth in that. But I did not have a voice when all this was done. And my intention, now that I am almost certain of being listened to, was, as soon as Mr. Galle had answered, to ask that the King's justice be extended to Mr. Encke. Nothing like this is possible as long as these people persist in their strange silence. I regret it a lot. The goodness of the relations between scientists is so nice but unfortunately so rare! So, my dear friend, if you learn something, and if I can make some step agreeable to these people, I count that you will inform me; I am sure that you know that I am above all this childish susceptibility. (Le Verrier to Schumacher. 7 January 1847)

The Schumacher letter to which Le Verrier refers does not seem to have survived, but we can guess that the source of the rumour was Humboldt. In a letter to Schumacher on 10 November 1846 (Bierman 1979, 85), Humboldt relayed a "sad, sinister secret" that Encke told him (Humboldt) that prevented the King awarding Galle the "Roten Alderorden IV. Klasse" for the telescopic discovery of Neptune (cited from Knobloch 2013:59). Encke, it should be said, supported Galle receiving a significant pay raise.

Possibly finding himself under pressure, Encke had to explain his motives to the finance minister in a letter. Encke's point was very simple (Knobloch 2013:59): Galle found Neptune because he had the Academy chart. There was a bit of luck involved, and anyone else could have discovered it. That was why he thought an increase in salary was better suited than a royal medal. Encke also pointed out that when Hencke discovered Astrea in December 1845, using another of the Academy charts, he was given a handsome increase of his pension, but no medal.

Another point that floated with rumours among German astronomers was the speculation that Galle was actually not very confident in Le Verrier's results and did not much care for the observations, as the Struve's gossip suggests. Nevertheless, as Dick (1986:255–256) pointed out, the picture of an excited young astronomer who, in spite of the views of his director, enthusiastically searched for the planet might not be completely true. This picture entered the story through Galle's obituary and Grosser's book, but Galle himself was careful in stating that he had: "a certain moral obligation to look it up at the position in question" (Galle, 1877:350). That might have been an expression of the fact that Galle shared Encke's doubts, as argued by Dick, or perhaps, a very good way to argue for the telescope time with his director.

In 1882, Galle repeated this situation in somewhat different words (Galle, 1882b:96): "Immediately after the reception I showed Encke the letter, who had previously been very averse to taking this matter into consideration, but now agreed, but left the execution of this request, which I personally had received, to myself."

This is consistent with what Encke wrote in his letter of congratulation to Le Verrier, published in *Centenaire:*

> I dare say that without the letter you have kindly addressed to Mr. Galle, the search would not have been made in Berlin. Reading with great interest your excellent Memoir (Comptes rendus, August 31), I thought I could wait until more details were published; I had it about eight days before September 23, and certainly would have started the search right away, if the conviction of the necessity that a planet should exist would have been strong enough […] However your letter specially inviting Mr. Galle to occupy himself with this research, it was not necessary at all to urge him to engage in it. (Centenaire, 1911:21)

Encke was quite clear that Le Verrier's idea to write to Galle, and not to him, was crucial for the observations to take place. This is consistent with Bruhns's biography of Encke, which stressed that Encke allowed his assistants and students to pursue their own research (e.g. Galle worked on Fraunhofer lines independently of Encke; Bruhns, 1869:191), and therefore implies that Galle actually wanted to do the search. In the same letter Encke also writes "*Il y a eu beaucoup de bonheur dans cette recherché*" (There was, nevertheless, a great deal of luck in the search; Lequeux, 2013:37). He referred there to the existence of the Bremiker's map, but as this chapter shows, there were many happy coincidences that resulted in the discovery of Neptune in Berlin.

**6.8 German contribution**

The English and the French might have abused Airy most savagely, but the Germans turned out to be Airy's strongest ally in the moment when tensions were at the highest. The international scandal around Neptune's discovery erupted when John Herschel (1846:1019) announced on 3 October, in a British literary magazine *The Athenaeum,* that "a young Cambridge mathematician, Mr. Adams" also predicted the location of Neptune. At a meeting of the French Academy of Sciences in Paris, Arago declared that Le Verrier gave him his permission to choose the name of the new planet, and he called it "*Planète de Le Verrier.*"

It thus must have been a great surprise when on 15 October Challis described his own search in Cambridge together with the announcement of his pre-discovery observations made of Neptune in August 1846, as well as the new (and much more correct) orbit of the planet calculated by Adams based on Cambridge and Berlin observations (Challis, 1846a:1069). Challis and Adams went further and suggested a name for the planet: "Oceanus." Proposing a name was a bold move by the Cambridge astronomers, perhaps incited by a letter from Arago published in "Our Weekly Gossip" column of *The Athenaeum* on 10 October. It stated that the discoverer of the planet, Galle

> …appears disposed to call the new planet Janus, from considerations borrowed from the hypothesis that it may be on the confines of our Solar System. M. Le Verrier, to whom belongs the right of naming it, does not agree to the significative of the name of Janus, but will consent to any other – Neptune, for instance – which would have the assent of astronomers. (Arago, 1846:1046)

Galle did not feel he had any claim to naming the planet, and dropped the issue waiting to see what the final verdict would be. When he published observations of Neptune, for example in early 1847, he called it "Le Verrier's planet" (*Leverrier'schen Planet*), a much more neutral form than "*Planète Le Verrier*". Other German astronomers were much more involved than Galle, and some made a crucial contribution to establish the name Neptune. Four persons deserve to be mentioned here: Encke, Gauss, Schumacher and Wilhelm Struve. Each of them had an important role in building the case for adopting the name "Neptune," instead of "Le Verrier," "Oceanus," "Minerva," "Hyperion," or "Janus," as they appeared in the press or astronomical literature (see Chap. 5).

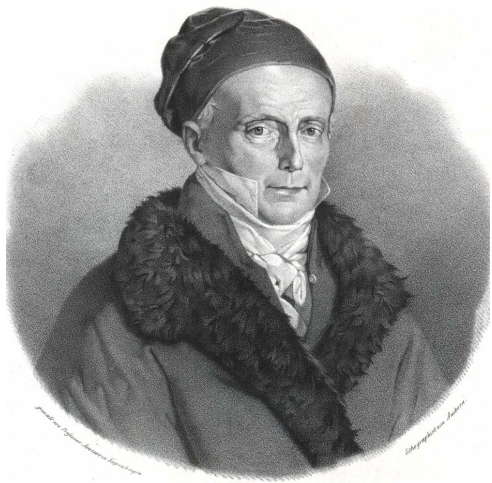

**Fig. 6.12** Heinrich Christian Schumacher (3 September 1780 – 28 December 1850) was a central figure in astronomical circles in Germany through his correspondence and his editing and publishing of the journal *Astronomische Nachrichten*. Lithograph by F. Ausborn based on a painting by Christian Albrecht Jensen (Wikipedia, Public Domain).

**6.8.1 Gauss and Schumacher at a "Possenspiel"**

Heinrich Christian Schumacher's (Fig. 6.12) mostly indirect contribution should be highlighted first. He was born in 1780 in Bramstedt (today called Bad Bramstedt) in province Holstein, then ruled by the King of Denmark, but part of the German Confederation. As a child he was introduced to the future king of Denmark, a meeting which helped secure a number of favours in later life. He finished a gymnasium in Altona and then studied in Kiel, Göttingen and received a doctorate in Dorpat (Tartu). Later he returned to Denmark, became a professor of astronomy at Copenhagen, but mostly lived in Altona where he founded an observatory that worked on the geodetic survey of Denmark.

His crucial contribution to astronomy was not with a discovery or scientific work in general, but by establishing, and editing until his death in 1850, the *Astronomische Nachrichten* (Astronomical News), the most important astronomy journal of the time. This journal, published continuously from 1821 to the present day, was truly international. It was different from other scientific journals of learned societies as it focused only on astronomy (including instrumentation and relevant mathematics), and, crucially, by its content it had a strong international presence. Astronomers across Europe, and later the world, sent letters with reports of their observations or calculations and Schumacher would publish them in the original form and their original language.

The impact of the *Astronomische Nachrichten* (Fig. 6.13) in the mid-1840s can be judged by the articles related to Neptune. The journal contains the report of the discovery (Encke 1846), the summary of Le Verrier's calculations (Le Verrier, 1846a, b, c), Challis's report on the Cambridge observations and Adams's new orbit (Challis, 1846b, c)[8], predating Challis's address at the Royal Astronomical Society, as well as the reprint of Airy's Royal Astronomical Society memorandum (Airy 1846b, 1847a). Reports on the observations of Neptune were numerous, particularly in Volume 25 of the journal, coming from observatories across Europe, extending from Russia to the British Isles and Ireland, as well as reports from the U.S. *Astronomische Nachrichten* also included Lassell's discovery of Triton (Lassell 1847), new calculations of the orbit by American astronomers (e.g., Henry and Walker 1847; Everett 1847; Pierce 1848), and was used as a sign of scientific quality (Hubbell and Smith, 1992) in the controversy of the "happy accident" (see Chap. 9). Volume 25 is also a record of astronomers' attitudes to the new planet's name.

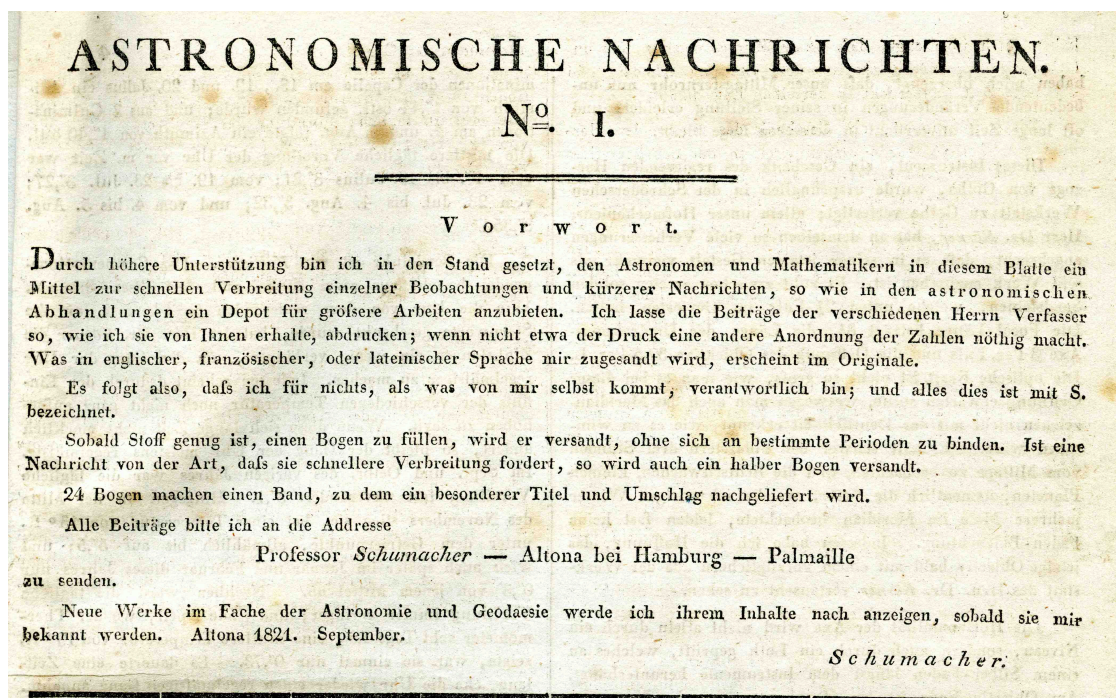

**Fig. 6.13** The first page of *Astronomische Nachrichten* in 1821 with Schumacher's mission statement for the new journal and the address to which contributions could be sent. Schumacher assures that the print will be as close as possible to the original, including the original language (if in English, French, or Latin, although some articles also appeared in Italian), and would be distributed as soon as there was enough material for printing. Schumacher was responsible for contributions signed "S." (Credit: Library of Leibniz-Institut für Astrophysik Potsdam).

As the editor of such a journal, Schumacher was in the middle of a vast correspondence network of astronomers with whom he exchanged information and gossip. Regarding the Neptune story, we have seen Schumacher's, perhaps very influential, advice to Le Verrier of where to look for astronomers with good and large telescopes. The naming issue featured prominently among the journal's correspondence. Arago, for example, asked him in a letter on 6 October (printed 22 October) to publish the official announcement of the planet's new name "Le Verrier" (Schumacher, 1846a). He also exchanged private letters with Le Verrier and Airy about the issue. The correspondence between Schumacher and Airy is particularly interesting for understanding how the name "Neptune" was adopted, and we will come to it later.

---

[8] This paper is somewhat more scientific version of the paper Challis published in *The Athenaeum* (Challis, 1846a), but it does not contain any mention of the Challis's and Adams's suggestion of the name "Oceanus." It seems that Challis initially wanted the suggested name to appear in the *Astronomische Nachrichten*, but Schumacher writes to Airy that it was omitted because Challis later requested to remove that part from the printed version (Schumacher to Airy, 11 December 1846).

Schumacher was on very friendly terms with the mathematician Johann Carl Friedrich Gauss, and they exchanged dozens of letters on the topic of Neptune, and specifically its name (Wattenberg, 1962). Almost all letters between Schumacher and Gauss in the period after the discovery, until the summer of 1847, at least partly dealt with the naming issue (Briefwechsel, 1863a, b). For example, when Gauss sent his first observations of Neptune, Schumacher delayed publishing them until he got an answer from Gauss about what he wanted to call the planet in print. In the beginning Gauss opted for the neutral "Leverrier's planet" (Gauss to Schumcher, 31.10.1846; Briefwechsel, 1863b:223-224), but soon after stated: "*Den Namen Le Verrier werde ich schwerlich jemals gebrauchen, weil ich es unschicklich finde*" (I will hardly ever use the name Le Verrier because I find it improper; Gauss to Schumcher, 31 October 1846; Briefwechsel, 1863b:224). At one point Gauss referred to the planet as "*Adams-Leverrier'schen Planet*" (Gauss to Schumacher 18 December; Briefwechsel, 1863b:265), probably the only instance it was called that.

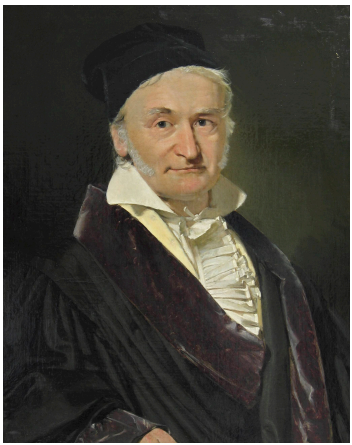

**Figure 6.14** Johann Carl Friedrich Gauss (30 April 1777 - 23 February 1855), oil painting from 1840 by Christian Albrecht Jensen. The leading mathematician of his time, Gauss was likely the only person to refer to Neptune as "Adams-Le Verrier's" planet (Wikipedia, Public Domain).

Gauss's and Schumacher's exchange of letters about the naming of the new planet focused on two themes of particular importance (Wattenberg, 1962). The first one was about the origin of the idea that Arago could name the planet. Schumacher, as a trained lawyer, pronounced that there was no legal problem as Le Verrier was present when Arago had announced that Le Verrier passed the right to him, and Le Verrier did not contradict this (Schumacher to Gauss 24 October 1846; Briefwechsel, 1863b:219).

Schumacher, however, suspected that this "transfer of right" was actually the work of Le Verrier, who could not the name the planet directly for himself. At first Gauss, could not believe that something like that could have really taken place, and referring to Arago he stated (Gauss to Schumacher, 31 October 1846; Briefwechsel, 1863b:223–224): "Incidentally, what you mentioned in your letter, and had earlier mentioned that L. V. had only advanced it to Ar[ago], seemed incredible to me then; I could not imagine that a man of real merit could have such a childish vanity."

His opinion was rather that the fault came from Arago "…who always likes to blatantly push himself forward, wanted to make himself important by such a protection of his protégé, while to him [Le Verrier] it must be extremely unpleasant, and he, as a subordinate of Arago (LV, as far as I know, is a kind of assistant to the Observatory) did not have the courage to contradict his patron."

Later Gauss changed his opinion and thought that Le Verrier himself had chosen the name "Neptune." He pointed to a newspaper article (Gauss to Schumacher, 12 December 1846 and 18 December; Briefwechsel, 1863b:259–260,265–267) from the Parisian newspaper *Galignani*

*Messanger*, published on 2 October (No. 9851), but Gauss read it on the sixth. The text, which Gauss cut out and saved, is identical to that published in *The Athenaeum* under the Gossip column on 3 October and sent by Arago (cited above). To Gauss this was a confirmation that the name "Neptune" was not chosen by the Bureau des longitudes, but by Le Verrier and therefore was appropriate to be used for the planet. The whole situation annoyed Gauss and he wrote to Schumacher (Gauss to Schumacher, 18 December 1846; Briefwechselb, 1863:265–267): "After this explanation was made, I can only consider the entire later comedy like a Possenspiel [a farce]. It means to take astronomers as fools, if you ask them today what name you want to choose, say Neptune, I give my approval in advance and tomorrow No, it should have my name!"

The second issue Schumacher and Gauss discussed enthusiastically in letters was what the name should actually be. Gauss first learned from Encke the suggestion of the name Neptune, together with the sign of a trident, appropriate ("*schicklich*"). Schumacher was not happy with that name, as it associated the planet's name with the sea. He could not think about Neptune (Schumacher to Gauss, 16 December 1846; Briefwechsel, 1863b:263–265): "… without being dragged by dolphins or other sea-beasts in a seashell-wagon on the sea, and that my imagination finds it too difficult to bring the sea to the sky."

The discussion between Schumacher and Gauss was often relaxed, reflecting their good relationship, and the letters were not necessary intended for anyone else to see. Notable was Schumacher's letter to Gauss where he compared the English search for Neptune to a contemporary detective story (Schumacher to Gauss 24 October 1846; Briefwechsel, 1863b:215 Schumacher 1846d): "If you have the Athenaeum of 17 October in Göttingen, you will see from [Challis] a proof of incomprehensible astronomical carelessness and a most admirable police report on the new planet, a little miniature piece of joke, as probably no one but de Morgan could have written."

After being informed by Thomas John Hussey that Airy is preparing a "pamphlet" about the discovery (Hussey refers to Airy's (1846a) account of the Neptune discovery), Schumacher wrote to Gauss (Schumacher to Gauss, 20 November 1846; Briefwechsel, 1863b:238): "God grant that a political war between England and France will not follow soon."

Gauss's contribution to the humorous side of the Neptune naming came in a letter written on 10 February 1847, where Gauss related a dinner joke that (Gauss to Schumacher, 10 February 1847; Briefwechsel, 1863b:283–284) "…the god who made the old daddy Uranus tumble, could be none other than Bacchus, and for the designation there is no more beautiful symbol than a wineglass, which at the same time will satisfy perfectly the ambitions of Le Verrier."

A few months later Gauss wrote (Gauss to Schumacher, no date; Briefwechsel, 1863b:330–331): "The naming of the transuranic planet is in all ways an unclean story.*"* Even more so, he continues "…if it is true that his [Le Verrier's] wife played the lead role in it." Elaborating it further, Gauss added:

> The joke of the Parisians, who call her Madame Thetis, has something mathematical about it: it is the counterpart to the conclusion:
> $$\text{If } x = y,$$
> $$\text{then } 1/2x = 1/2y.$$
> Especially since
> $$\text{Le Verrier} = \text{Neptune},$$
> so is
> $$\text{the (marriage) half of LeVerrier} = \text{half of Neptune}.$$

Schumacher did not want to be outplayed by Gauss and responded (Schumacher to Gauss, 26 July 1847; Briefwechsel 1863b:332): "That Madame Thetis leads the regiment, I have also heard, so it is here $1/2x \sim x$, or Hesiod's statement "δτι πλεον ημισυ παντος" [because it's halfway through] arrives. From the previous inequality would follow $0 > 1/2x$, which is only possible in the not so flattering assumption for [Le Verrier] that x is a negative quantity."

To this position of joking at the expense of Le Verrier, Gauss and Schumacher arrived over a course of several months, in which they convinced themselves that Le Verrier was playing a double game both with Arago and the astronomical community. Schumacher, based on correspondence with Le Verrier, wrote (Schumacher to Gauss, 23 March 1847; Briefwechsel, 1863b:286–287): "Arago's great anger at the word "confus" [my quotes, for more details see 6.8.2] seems to me hard to understand, if one does not want to presume that the idea of the Cession of the Right, to thereby keep the name Le Verrier, had emanated from Le Verrier himself, and that Arago believed Le Verrier wanted through his letter to Encke to shove everything on Arago."

**6.8.2 The German Choice**

Of all the different ways German astronomers named the planet in 1846 and early 1847 publications, the most common was the neutral description "Le Verrier's planet." Privately there were other suggestions; for example, Encke considered "Vulcan" (Holland, 1872:299). Humboldt suggested "Erebus" in a letter to Schumacher (Wattenberg 1962:8), but in a letter to Galle he called it "Poseidon" (Watenberg, 1963:53), perhaps as a joke.

The opinion of German astronomers was summarised by Schumacher in a letter to Airy at the end of 1846. It was first Airy who in a letter to Schumacher (Airy to Schumacher, 9 December 1846) asked: "I would be glad to know the opinion of yourself and the Germans on the planet's name. I think that both Germans and English have a right to be heard in this matter."

Airy also noted why he disliked the name "Neptune" and mentioned John Herschel's suggestion "Minerva," as well as other English suggestions such as "Oceanus" and "Hyperion." Schumacher's reply was split among four different letters Schumacher wrote to Airy in December 1846, but the one of 18 December was most explicit. Schumacher started his explanation of the German opinion: "My German friends find Arago's arrogance intolerable" (Schumacher to Airy, 18 December 1846). He explained that the German astronomers could not agree with Arago's dismissal of Adams and his contribution:

> The priority depends upon the publication and neither you, nor we contest Mr Le Verrier's claim to legal priority, but we agree with you that there is no doubt, that Mr Adams has found the place of the new Planet, before Mr Le Verrier. We think at the same time that Mr Le Verrier's research were deeper and rested on more solid ground than those of Mr Adams, as far as we know the last. My German friends incline generally for the name of Neptune (which disagrees with me, as with you) because they suppose that this was the name, to which on the proposal of the Board of Longitude Mr Le Verrier gave his consent, before Mr Arago persuaded him to delegate his right to the Secrétaire perpétual, in order to have named the Planet LeVerrier [underscored and written as a one word].

Schumacher's private letter might have influenced Airy's thinking, but it was however, the publicly announced choices of Encke and W. Struve that helped fix the name "Neptune."

On 22 October 1846, Encke presented to the Royal Academy of Berlin the news of the discovery of the planet and devoted a large fraction of his paper to the question of the planet's name (Encke, 1847). He also sent the paper to Schumacher in December 1846 and it was duly printed on 11 February 1847, in No. 588 of the *Astronomische Nachrichten*. Like many other observatory directors who were publishing astronomical almanacs, Encke had to decide what to call the new planet. Aware of the raging controversy, he used the presentation as a way to argue for the name he liked the most.

He started with a historical overview, mentioning both the naming practices for Uranus and Ceres. In Encke's view, Uranus originally was called the "Georgian planet," as gratitude to King George III, who "enabled [William] Herschel to build his large reflector". Ceres was originally called "Fedinandea, to pay tribute to the King of Naples, the founder of the Palermo observatory," Those names did not endure and by the 1840s they were universally called "Uranus" (except, as Encke remarked, in England) and "Ceres." Encke pointed out that it was Bode who suggested the name "Uranus," and with the German discoveries of minor planets by Harding, Olbers (two) and the most recent by Hencke, all those bodies were given mythological names. He also made the point that Pallas was sometimes referred to as "Olbersiana," until Olbers declared himself strongly against it.

Encke continued to build his case by quoting from the letter Le Verrier sent to Galle on 1 October 1846 (see Chap. 5 and Chap. 7): "The Board of Longitudes here has proposed the name of Neptune, with a trident as symbol. The name of Janus would mean that this planet is the last of the Solar System, something that there is no reason to believe. "

## 6. "That star is not on the map": The German Side of the Discovery

As the final point in his case, he quoted a letter from Gauss, *"unserer ersten deutschen astronomischen Autorität"* (our first German astronomical authority): "I find the name Neptune chosen by Mr. Le Verrier perfectly decent: as a sign one could perhaps choose a trident, if it were not inappropriate to anticipate the author in any way."

Encke, who, like anyone else, did not know if Le Verrier actually chose Neptune (or if it was chosen by the Bureau des Longitudes), used Gauss's statement to start building the case that Le Verrier was not against that particular name, in contrast to Arago's proposal.

To understand Encke's conviction we should look at what Sir Henry Holland (1788 – 1873), a royal physician and avid traveller, wrote in his autobiography about a night he spent at Berlin Observatory, some "ten or twelve days after the discovery". Holland heard about the discovery while visiting Bremen and hurried to Berlin. As it was a cloudy night they couldn't observe, but he reports the following scene:

> Among other things discussed while thus sitting together in a sort of tremulous impatience, was the name to be given to the new planet. Encke told me he had thought of Vulcan; but deemed it right to remit the choice to Leverrier, then supposed the sole indicator of the planet and its place in the heavens—adding that he expected Leverrier's answer by the first post. Not an hour had elapsed before a knock at the door of the Observatory announced the letter expected. Encke read it aloud; and, coming to the passage where Leverrier proposed the name of Neptune, exclaimed, "*So lass den Namen Neptun sein.*" It was a midnight scene not easily to be forgotten. A royal baptism, with its long array of titles, would ill compare with this simple naming of the remote and solitary planet thus wonderfully discovered. There is no place, indeed, where the grandeur and wild ambitions of the world are so thoroughly rebuked, and dwarfed into littleness, as in the Astronomical Observatory. (Holland, 1872:299)

The letter Holland mentioned was most likely the letter from Le Verrier to Galle, which arrived at Berlin at the beginning of October (consistent with Holland's visit to the Observatory) rejecting Galle's name suggestion (Le Verrier to Galle, 1 October 1846).

As the final point in his argument for choosing the name "Neptune," Encke presented an extract from Le Verrier's letter sent to him on 6 October, a day after Arago announced his choice for the name. We present it here in the French original, as, according to Le Verrier's later statements, the meaning of the sentence was misunderstood:

> *J'ai prié mon illustre ami Mr. Arago de se charger de ce soin. J'ai été un peu confus de la décision qu'il a prise dans le sein de l'Académie.* (I asked my illustrious friend Mr. Arago to take in charge the choice of a name for the planet. I have been somewhat confused by the decision he has taken at the Academy.)

The printing of this sentence, without Le Verrier's approval, contributed significantly to the collapsing relations between Le Verrier and Arago in Paris, as Le Verrier reported to Schumacher. From the correspondence between Le Verrier and Schumacher it is evident that Le Verrier was annoyed that Encke printed it. His point was that it was printed out of context and that the word "confus" was translated incorrectly (Le Verrier to Schumacher 16 March 1847). This was what Le Verrier wrote to Schumacher to explain the tense situation in which he found himself in Paris:

> One was searching for a long time a pretext to attack me. You will probably judge, Monsieur and friend, that this was not easy, since one had to be content with the following pretext. Mr. Encke published incautiously a few lines of a letter I wrote him, the meaning of it he has not understood. I rewrite the passage entirely now. You will see that it was only question of explaining to Mr. Encke what Mr. Arago has done, and not at all to disavow him [Arago]. You should remember that we say to someone to whom we owe much: Mr., I am confused ["embarrassed" could be a more suited translation of the world "confus" in this context] of your kindness. This is the meaning of the word "confus".

> Mr. Schumacher made me the honour to write to me, looking for a name for the planet. I asked my illustrious friend Mr. Arago to take in charge the choice of a name for the planet. I

> have been somewhat confused [embarrassed] by the decision he has taken at the Academy. I would not have dared to explain you what this decision is if I had not found in it the opportunity to pay a just tribute of admiration for your work on Encke's comet. The obscure name that Mr Arago wants to give to the Planet would put me on the same glorious footing as the illustrious director of the Berlin Observatory, and I do not deserve this.

The problem stemmed from misunderstanding the French word "confus". According to Le Verrier, the true meaning of the sentence published by Encke was related to the next sentences in the letter (not used by Encke): Le Verrier was embarrassed, in a flattered way, that his name would be compared to the name of such a famous person such as Encke. It is quite clear, however, that Encke would have not wished to print such a flattering sentence, and perhaps Le Verrier was correct to complain that his statement had been taken out of context.

Whether or not Le Verrier had a secret agenda when writing to Encke, as suspected by Gauss and Schumacher, and regardless of the real meaning of the word, Encke assumed that Le Verrier himself did not really approve of Arago's choice. Therefore, Encke concluded: "Under these circumstances I will, supported by the high authorities of the Bureau des Longitudes in Paris, and of Geh. Hofraths Gauss, keep the name Neptune and the sign of the trident for the next few years, until public opinion in Germany is sufficiently consolidated to establish a definite name."

The paper printed in the *Astronomische Nachrichten* finished with a paragraph that looks like an add-on to the original presentation at the Academy. Encke stated that in "later letters" (that arrived after his presentation), the name "Neptune" was accepted by Sir John Herschel and Wilhelm Struve. The last sentence also claimed that "the first astronomical authorities in Germany, France, England and Russia" adopted the name (Encke, 1847:196).

The letter from Wilhelm Struve, the director of the Pulkovo Observatory, to which Encke referred, must have been similar to the letter Wilhelm Struve sent to Airy (Struve to Airy, 23 January 1847), which included a copy of the essay published by Struve (Struve, 1848) in the Bulletin of the St. Petersburg Imperial Academy of Science. Airy re-published it in *The Athenaeum* on 20 February 1847 (Airy, 1847b:199). W. Struve's essay was titled: "On the denomination of the planet newly discovered beyond the orbit of Uranus". It presented a list of arguments why the name "Neptune" was appropriate and why Arago's suggestion of "Le Verrier" was not.

W. Struve started from the same problem as Encke, what name to give to the new planet in their yearly publications with ephemerides? A number of arguments were similar, such as the historical parallels with Uranus, including that even John Herschel called it "Uranus," and the fact that the Bureau des Longitudes proposed "Neptune" and Le Verrier did not raise any objections to it. Struve's arguments against "Le Verrier" were also similar: all planets should have mythological names and not names of discoverers.

More remarkable is his second argument against the name, as it was not only Le Verrier who predicted the location of the planet, but also John Couch Adams, and that had to be taken into account. W. Struve concluded his letter to Airy in a similar way as Encke (Airy, 1847b:199): "Consequently, we will retain the name of Neptune; and will make no change, unless hereafter the general voice shall determine in favour of another name."

These two powerful statements by the most prominent of German astronomers, that appeared at similar time in print, in two scientific journals, and in a literary magazine, read by the educated people across Europe, finally settled the controversy over the name of the new planet.

The Neptune naming scandal that erupted almost immediately after the announcement of the discovery bears witness that science is a human endeavour after all. This is also the most amusing part of the story, especially when one considers some of the semi-private opinions voiced by renowned scientists and respected members of society. Gauss's and Schumacher's remarks are just some that are available to us, as they have been preserved in their letters (and even published in books celebrating their achievements.) Another infamous, but still entertaining, remark on the topic came from a letter to Airy sent by Captain William Henry Smyth (1788 – 1865) on 5 December 1846, then president of the Royal Astronomical Society:

> I don't quite like this proposed change in the nomenclature of the Planets, for mythology is neutral ground. Herschel is a good name enough. Le Verrier somehow or other suggests the idea of a Fabriquant & is therefore not so good. But just think how awkward it would be if the next planet should be discovered by a German: by a Bugge, a Funk, or your hirsuite friend Boguslawski! (Smyth to Airy, 5 December 1846)

Captain Smyth's letter was written before the publication of Encke's paper (Encke, 1847), in which Encke did not fail to mention that "Our German custom has established itself at four, you can say, at five new planets, since Herschel is a German by birth ..."

Given the development of astronomy within German-speaking countries, and the efforts described in this chapter, it is perhaps not so exceptional to consider that by 1846 of seven new planetary bodies, five were discovered by Germans: Uranus (Herschel in Bath, England), Pallas (Olbers in Bremen), Juno (Harding in Lilienthal), Vesta (Olbers in Bremen), and Astrea (Hencke in Driessen). Furthermore, Ceres was re-discovered by a German (Olbers) based by the prediction made by a German (Gauss). Finally, Neptune *was* discovered in Berlin (Galle and d'Arrest), based on the prediction by Le Verrier.

**Acknowledgements.** This chapter would not be possible without the help from the librarians of the Leibniz-Institut für Astrophysik-Potsdam (AIP), Regina von Berlepsch and Melissa Thies. Thank you for your indefatigable and very resourceful support. A special thanks goes to the director of AIP, Matthias Steinmetz, who introduced me to the intricate details of the Neptune discovery some years ago over a good dinner in Lyon. I would also like to thank Wolfgang Dick for pointing me to the location of Galle's letters and useful discussions. Sabine Thater helped with translations from German originals, while Damien Le Borgne for initial and James Lequeux for final translations of Le Verrier's and Galle's letters from the French. James, thank you for being so responsive. I had several stimulating email exchanges with a number of co-writers of this book, which I thoroughly enjoyed. I would like to thank Brian Sheen for inviting me to join the project, and the editors William Sheehan, Robert W. Smith, Carolyn Kennett and Trudy E. Bell for many inspiring thoughts and patience in waiting for my contribution.